\documentclass{article}
\usepackage{arxiv}

 \usepackage{epsfig}
\usepackage{amsmath}
\usepackage{amssymb}
\usepackage{amsthm}
\usepackage{amsfonts}
\usepackage{latexsym}
\usepackage{graphicx}
\usepackage{caption}
\usepackage{enumitem}
\usepackage{multirow}
\usepackage[usenames,dvipsnames]{xcolor}
\usepackage{color, colortbl}
\usepackage{bbm}
\usepackage{algorithm}
\usepackage{algpseudocode}
\usepackage{url}
\usepackage{pifont}
\usepackage{geometry}
\usepackage{dirtytalk}
\usepackage{footnote}
\usepackage[many]{tcolorbox}
\usepackage[section]{placeins}
\usepackage{mathtools}
\usepackage{tfrupee}
\usepackage{subcaption}
\usepackage{framed}
\usepackage[colorlinks=true,linkcolor=blue,citecolor=black]{hyperref}
\usepackage{mathrsfs}
\usepackage{varwidth}
\usepackage{booktabs}
\usepackage[
maxbibnames=99,
backend=biber,
style=numeric,
]{biblatex}
\usepackage[
backend=biber,
style=numeric,
]{biblatex}
\newcommand{\citet}[1]{\citeauthor*{#1}~\citeyear*{#1}~\cite{#1}}
\newcommand{\citep}[1]{\cite{#1}}

\newcommand{\myemph}[1]{{\em \textcolor{blue}{#1}}}

\graphicspath{{images/}{./}} 

\newtheorem{theorem}{Theorem}[section]

\newtheorem{example}{Example}[section]

\newtheorem{definition}{Definition}[section]

\newtheorem{remark}{Remark}[section]

\usepackage{graphicx}
\usepackage{amssymb}
\usepackage{amsmath}
\usepackage{xcolor}
\usepackage{soul}
\usepackage{enumitem}
\usepackage{centernot}
\usepackage{enumitem}
\usepackage{multirow}
\usepackage{booktabs}
\usepackage{mathtools}
\usepackage{booktabs}
\usepackage{longtable}
\usepackage{tabu}
\usepackage{lscape}
\usepackage{geometry}
\usepackage{rotating}
\usepackage{breakcites}
\usepackage{adjustbox}

\long\def\symbolfootnote[#1]#2{\begingroup%
\def\thefootnote{\fnsymbol{footnote}}\footnote[#1]{#2}\endgroup}

\onecolumn
\usepackage{ragged2e}
\usepackage[many]{tcolorbox}

\title{Fair Allocation of goods and chores -- Tutorial and Survey of Recent Results}

\author{
  Shaily Mishra, Manisha Padala, Sujit Gujar \\
  Machine Learning Lab\\
  International Institute of Information Technology (IIIT), Hyderabad \\
  \texttt{\{shaily.mishra, manisha.padala\}@research.iiit.ac.in} \\
  \texttt{sujit.gujar@iiit.ac.in}
  }

\date{July 2023}

\begin{document}

\maketitle

\begin{abstract}
Fair resource allocation is an important problem in many real-world scenarios, where resources such as goods and chores must be allocated among agents. 
In this survey, we delve into the intricacies of fair allocation, focusing specifically on the challenges associated with indivisible resources.
We define fairness and efficiency within this context and thoroughly survey existential results, algorithms, and approximations that satisfy various fairness criteria, including envyfreeness, proportionality, MMS, and their relaxations. Additionally, we discuss algorithms that achieve fairness and efficiency, such as Pareto Optimality and Utilitarian Welfare.
We also study the computational complexity of these algorithms, the likelihood of finding fair allocations, and the price of fairness for each fairness notion. We also cover mixed instances of indivisible and divisible items and investigate different valuation and allocation settings.
By summarizing the state-of-the-art research, this survey provides valuable insights into fair resource allocation of indivisible goods and chores, highlighting computational complexities, fairness guarantees, and trade-offs between fairness and efficiency. It serves as a foundation for future advancements in this vital field.

\end{abstract}


\section{Introduction}

\noindent Fair division of resources is a well-studied problem in the division of inheritance, divorce settlement, treaty negotiations, and resolving disputes.  Polish mathematicians Steinhaus, Banach, and Knaster~\cite{propSTEIHAUS} are credited with being the first to study this problem formally. Since then, the topic has piqued interest both in social sciences and game theory, garnering various academic books written on this topic like \citet{brams_taylor_1996}, \citet{moulinfairness}, \citet{webbcakecutting}. Most of the existing literature focuses on resources that can be divided, such as land or money. In these cases, the concept of fair distribution is often explored through the lens of cake cutting, with the classic ``divide and choose'' algorithm as the cornerstone. Many intriguing notions of fairness and corresponding algorithms have been proposed within this framework. However, when resources are indivisible, it introduces even more captivating challenges to the field.

In the real world, numerous examples of fair division have emerged to address allocation challenges across various domains. Notable instances include Course Match (coursematch.io), Spliddit (spliddit.org), and Fair Outcomes (fairoutcomes.com) offer practical applications of fair allocation algorithms for diverse scenarios.
These algorithms cater to allocation dilemmas such as course allocation, equitable rent distribution among roommates, sharing taxi fares, and assigning goods among a group of individuals. Moreover, allocating chores is equally important, especially for gig economies. For example, food delivery platforms worry about distributing delivery orders among the available drivers, and other platforms like Uber worry about distributing rides. The gig economies have the responsibility of maintaining and ensuring the customers are happy by reducing delivery/travel times while making sure the employees are not over or underworked. Hence both efficiency and fairness matter. More importantly, with the advent of the internet, large tech companies have pushed towards distributed computing. Allocating multiple resources to multiple available tasks to distribute the task load and facilitate maximum utilisation of available resources. Due to the rapid advancement in technology, such as Artificial Intelligence (AI) and the Internet of Things (IoT), many autonomous agents encounter the challenge of task or resource allocation. Fairness in resource allocation is essential for AI's sustainable and responsible growth and for minimizing biases and discrimination.

Given the limited availability of resources, addressing fairness and efficiency concerns during the allocation process is crucial. While fairness ensures that each individual receives their fair share based on relevant criteria, efficiency aims to optimize the allocation to maximize the benefit of the limited resources. Therefore, achieving fairness and efficiency is of utmost importance when allocating limited resources among diverse groups of individuals. In this setting, we explore several questions. Firstly, we define what constitutes a fair allocation and whether such allocations exist. If they do, how can we compute them? Alternatively, if fair allocations do not exist, can we relax our fairness criteria?
Once we have achieved a fair allocation, can we obtain an allocation that maximizes efficiency among the set of fair allocations? How can we achieve both fairness and efficiency in our allocations? This survey addresses fundamental questions concerning attaining \myemph{fair and efficient allocations}. 

Extensive research has been undertaken in the realm of fair and efficient allocation of goods and chores. In this survey, we aim to provide a comprehensive tutorial-like approach, delving into intricate details concerning allocating goods, chores, and their combined scenarios, along with various valuations and discussing the latest findings and algorithms about these allocation domains.

Numerous surveys have delved into this subject, offering valuable insights. For instance,~\cite{amanatidis2023fair} examines the recent advancements in dividing indivisible goods, presenting results and algorithms, \cite{aziz2022algorithmic} provides a short survey on the algorithmic aspects of fairness notions of indivisible items, exploring open questions in the field. These surveys serve as illuminating resources for understanding the current state and future directions of fair allocation,~\cite{aziz2020developments} provides a survey of fair division in multi-agents,~\cite{aleksandrov2020onlinesurvey} focuses on fair division in online settings. For other results, there are several surveys shedding light on fair allocations~\cite{moulin2019fairinternetage,suksompong2021constraints,bouveret2016fair}. 
However, our survey is recent with a broader scope. Most surveys focus solely on goods or chores; we have diligently combined both aspects, providing a comprehensive analysis. In addition, we review results for a combination of goods and chores and delve into situations where there is a mixture of divisible and indivisible resources. In summary, our survey focuses on facilitating comprehension and comparison of the surveyed results; we present a comprehensive summary in Table~\ref{tab:summarytable1}. This table is valuable for researchers and practitioners, providing a concise overview of the surveyed findings, algorithms, and critical insights. Additionally, the survey also aims to provide a tutorial on fair division -- explaining the various fairness and efficiency notions with examples. It lets readers quickly understand the state-of-the-art advancements in fair and efficient allocation without any prior knowledge in the domain. In addition to focusing on indivisible goods and chores, we briefly survey the fair division literature concerning divisible and indivisible items. By exploring both domains, we aim to provide a more holistic understanding of fair allocation mechanisms across a broader range of resource types. Again, we summarize the key findings and approaches in Table~\ref{tab:mixedmodelresults}, offering a convenient reference for readers interested in the fair division of divisible and indivisible items. Overall, our survey goes beyond a mere compilation of existing research. It provides a tutorial-like exposition, offering detailed insights, algorithmic approaches, and open questions regarding fair and efficient allocation. The summarized tables enhance the accessibility of the surveyed information, facilitating further research and advancements in this vital domain.

\smallskip
\noindent{{\myemph{Organization of this survey}}} In Section~\ref{sec:motivation}, first, we explain the need for fair division through a few examples.
In Section~\ref{sec:prelims}, we explain important definitions related to fairness and efficiency. Building upon this foundation, we then delve into the topic of fair and efficient allocation in the context of indivisible goods in Section~\ref{sec:survey_goods}. Subsequently, we explore the intricacies of fair and efficient allocation in the domain of indivisible chores in Section~\ref{sec:chores}. To further enhance our understanding, we investigate the combined setting of indivisible goods and chores in Section~\ref{sec:combination}. This section explores the challenges and considerations when allocating resources that encompass goods and chores. In Section~\ref{sec:mixedmodel}, we focus on the mixed setting of indivisible and divisible items. Here, we examine the implications and approaches of fair allocation when resources include both items. Continuing our exploration, Section~\ref{sec:otherresults} analyzes various results related to the likelihood of fair allocation, the price of fairness, and different fairness settings. This section highlights the probability of achieving fair allocations, the price of fairness, and the different perspectives on defining fairness. Finally, we conclude our survey in Section~\ref{sec:sum}, summarizing the key insights and open questions that emerged from examining fair and efficient resource allocation in the context of indivisible goods and chores.

\section{Motivation and the Problem}
\label{sec:motivation}

We will illustrate the problem of fair allocation, starting with a simple example.

\begin{tcolorbox}[enhanced,drop fuzzy shadow southwest, colframe=blue!30!white!20!black,colback=blue!10]\begin{example}
\label{ex:cakeexample}
Two intelligent friends, Agastya and Noorie, won a contest and received an uncut veggie pizza as a reward. These friends need to divide this pizza fairly.
\end{example}
\end{tcolorbox}

Suppose both of them like veggie pizza equally; hence the fairest way to divide it is to cut it equally. Let's say that the pizza has various toppings, such as pineapple, mushroom, olives, and corn. If both like all the toppings, they can cut the pizza equally for each topping. However, what if they have their preferences? Imagine Agastya dislikes pineapple toppings, and Noorie is allergic to mushrooms. Can they cut the pizza according to their preferences?

Noorie suggests splitting the pizza in two, and Agastya could choose first what he likes while she takes the remaining piece. Would this be fair? Let's analyze this approach. Noorie cuts the pizza so that both the partitions are equal to her. As a result, she is happy with any piece she gets. Agastya picks the pizza part he likes the most. This division is famously known as the \myemph{Cut-and-Choose algorithm, `I cut, you choose'}.

The above pizza was uncut, and the agents were free to cut the pizza in any way they liked. These types of items are called \myemph{divisible items}.

\begin{tcolorbox}[enhanced,drop fuzzy shadow southwest,colframe=blue!30!white!20!black,colback=blue!10]
\begin{example}
\label{ex:indivisibleexample}
Let us, however, consider a modification where Agastya and Noorie receive eight pieces of pizza and cannot cut the pieces any further. Either they take the whole piece, or they don't. They both like every piece equally. This setting is called as \myemph{indivisible items}. Each person can take four pieces, so the allocations are fair. Consider, however, if they only received seven pieces. Is it possible to distribute the pieces fairly among them in such a case?  
\end{example}
\end{tcolorbox}

Before we can answer this, let's look at the following situation.

\begin{tcolorbox}[enhanced,drop fuzzy shadow southwest, colframe=blue!30!white!20!black,colback=blue!10]
\begin{example}
\label{ex:choreexample}
Agastya and Noorie need to clean their house. This includes tasks such as cleaning the living room, clearing the kitchen mess, dusting the furniture, etc. Having fewer chores is what everyone wants. What is the best way to divide the tasks such that everyone gets their fair share?
\end{example}
\end{tcolorbox}


Comparing Example~\ref{ex:cakeexample} to Example~\ref{ex:choreexample}, in the first one, everyone would like to obtain a bigger share, while in the second, everyone is more interested in obtaining a smaller share. It is commonly known (in the literature) that items (resources/tasks) with such positive value are called \myemph{goods}, and those with negative value are called \myemph{chores}. The approach to solving fair division for goods and chores is quite different, even though it may seem like just a matter of negation. We will examine this in more detail later.

Suppose there are multiple friends -- \myemph{agents} in the system. How should we generalize this notion of \myemph{fairness}? The idea that agents should be happy with their allocations (allocated bundles) is \myemph{envy-freeness}~\cite{Foley1967ResourceAA}. Envy-freeness means agents/players value their bundle more than others, i.e., no agent envies others. We can always divide divisible items without envy. Allocating indivisible items without inciting envy is, however, a challenging task. For example, if there is only one good, and Noorie receives it, Agastya will envy her. It is widely known that EF is not guaranteed to exist for indivisible items. 

With the negative results concerning the existence of envy-freeness for indivisible items, the question naturally arises, can we ensure envy-freeness approximately, i.e., to relax the fairness notion? For Example~\ref{ex:indivisibleexample}, giving three of the seven pieces to one person and the remaining four to another is arguably the fairest way to allocate them. Assuming Agastya receives three pieces and Noorie receives four, he envies Noorie by at most one piece. Budish introduced such a relaxation of envy-freeness, called \myemph{Envy-freeness up to 1 item (EF1)}~\cite{budish2011EF1}. We will formally define all the fairness notions in detail later on, but in essence, EF1 implies that if envy exists in the system, an agent can only envy the other agent by at most one item. 


Finding a fair allocation in a simple example involving two friends was challenging. Here, we present a highly complex example to illustrate the problem's difficulty. Imagine that the Government wishes to invest in improving national security, economy, agriculture, and healthcare. It picks four re-owned scientists in these individuals' domains and asks them to set up a Research Lab for the nation's improvement. It provides them with a wide variety of resources such as professors, laboratory assistants, engineers, vehicles, books, subscriptions, journals, drones, microscopes, biometrics systems, computer servers, land, water facility, electricity facility, supercomputers, etc. This intricate scenario involves distributing many divisible and indivisible goods and responsibilities to the four scientists forming Research Centers. The Government wants to ensure that all four centers feel equal so that each one gives its best to develop this project. In addition, it wants to ensure that the resources are allocated to the Lab, which will put them to best use; for example, microscopes will benefit the pathology department in the Healthcare Center rather than the Economic Center. There are divisible items such as Land, Electric facilities, Water facilities, etc., and indivisible items such as Computer Systems, Servers, Books, etc. An item may be beneficial to one person but a chore to another. Microscopes, for example, may be valued positively by the Healthcare Center but negatively by the Economic Center since they are of no use to them. We should also strive to ensure \myemph{efficiency} among fair allocations; since, otherwise, not distributing any item to an agent is fair. We are looking for a \myemph{fair and efficient} allocation. 



Fairness depends heavily on the context; what may be fair in one setting may not be fair in another. Suppose Agastya and Noorie are supposed to purchase a new refrigerator. Agastya likes Samsung, and Noorie likes LG. Regardless of what kind of refrigerator they buy, the concept of envy is no longer applicable. We choose fairness criteria depending on the context. Later, we will discuss various kinds of fairness notions in detail. In essence, we will aim to formulate broadly acceptable fairness criteria. In doing so, we seek to determine whether it always exists. If it does, can we find it in a reasonable amount of time? In essence, we will aim to formulate broadly acceptable fairness criteria. In doing so, we seek to determine whether it always exists. If it does, can we find it in a reasonable amount of time? Is it possible to relax the fairness criteria to apply to the setting if one does not exist?


It is an \myemph{NP-complete} problem to determine whether there is an envy-free allocation when there are indivisible items, even when the valuation is binary~\cite{aziz2015fairordinal,hosseini2020fair}. Even if it exists, finding it is \myemph{NP-hard}~\cite{JAIR3213}. It is natural to look for fairness relaxations that exist as well as polynomial-time algorithms. Researchers are interested in studying relaxed fairness notions in light of the impossibility of fairness for indivisible items. Relaxations can be categorized into two: additive relaxations and multiplicative relaxations. Ideally, if we cannot find a fair distribution, we would prefer an allocation that approximates the fairness guarantees in some multiplicative way. Suppose, for example, that we cannot find a fair share allocation; we're aiming to find an allocation that approximates the fair share in some multiplicative way for each agent. However, fairness is difficult to achieve even with a multiplicative approximation. Therefore, we turn to additive approximation, which exhibits possibilities and impossibilities.

\section{Preliminaries}
\label{sec:prelims}
First, we define all essential concepts pertaining to fairness and introduce all needed notation.

\subsection{Fair Resource Allocation: Basic Model and Notations}

We consider the problem $\langle N,M,V \rangle$ of allocating $M$ indivisible resources/items among $N$ interested agents with valuation profile $v \in V$, where $M=[m]$, i.e., $M=\{1,2,\ldots,m\}$, $N=[n]$ and $m,n \in \mathbb{N}$. We define allocation $A\in \Pi_n(M)$ as the partition of $M$ indivisible items into $n$ bundles, i.e., $A=(A_1,A_2,\ldots,A_n)$, where $A_i$ is the bundle allocated to agent $i$. We only allow complete allocation, and no two agents can receive the same item. That is, $A=(A_1,A_2,\ldots, A_n)$, s.t., $\forall i, j \in N$, $i \neq j; A_i \cap A_j = \emptyset$ and $\bigcup_i A_i = M$. We denote the allocation for all the agents except $i$ as $A_{-i}$. An allocation is called partial if $\bigcup_i A_i \subset M$.

\subsubsection*{\myemph{\textbf{Goods vs Chores}}}
Each agent $i \in N$ has a valuation function $v_i : 2^{M} \rightarrow \mathbb{R}$, and $v_i(S)$ represents its valuation for the subset $S \subseteq M$. We represent the valuation profile as $v=(v_1,v_2,\ldots,v_n)$. For all agents, the valuations are normalized, i.e., $v_i(\emptyset) = 0$. We denote the valuation of an item $k\in M$ for any agent $i\in N$ as $v_i(\{k\})$ or $v_{ik}$. For an agent $i$, an item $k \in M$ is a \emph{good} if, $v_{ik}\geq 0$, and a \emph{chore} if, $v_{ik} < 0$. Goods are desirable/positively valued and chores are undesirable/negatively valued items. There could be three settings - pure goods, pure chores, and a combination of goods and chores. In combination, $\forall k \in M$ it is possible that, for some agent $i\in N$, $v_{ik}>0$ and for another agent $j\in N$, $v_{jk}<0$. I.e., an item may be good for one agent and a chore for another. Note that we use the terms -- valuation and utility interchangeably in this chapter.

\subsubsection*{\myemph{\textbf{Different Types of Valuation Functions}}}

In most of the literature, we assume that the valuation functions of agents are \emph{monotonic} for goods and \emph{anti-monotonic} for chores. If we increase goods in an agent's bundle, the utility increases, and if we increase chores, the utility decreases. Formally, we define it as follows,

\begin{definition}[\myemph{Monotonicity}]
\label{def:monotonicity}
A valuation function  $v_i$ is monotonic if, $\forall S\subseteq T \subseteq M, v_i(S) \le v_i(T)$.
\end{definition}

\begin{definition}[\myemph{Anti-monotonicity}]
\label{def:anti-mono}
A valuation function  $v_i$ is anti-monotonic if, $\forall S\subseteq T \subseteq M, v_i(S) \ge v_i(T)$.
\end{definition}

We say that agents have \emph{additive valuations} when, for each agent, the valuation of a non-empty bundle equals the sum of the valuations of the individual items. However, it is also possible that agents value an allocation more or less than the sum of the values of the individual items, i.e., super-additive or sub-additive. Furthermore, agents may have marginal decreases or increases in their valuation with every additional item, i.e., sub-modular, super-modular valuations.

In the most general setting, the agents may have arbitrary valuations for every subset i.e., the \emph{general valuation} setting. General valuations are rich and complex functions. Consider dividing 30 items among several agents. To provide a complete valuation of items, each agent should provide their value for every bundle of the items, i.e., they would have to report the value of more than 1 billion bundles. Hence in literature additive valuations are widely studied.

\begin{definition}[\myemph{Additive Valuations}]
\label{def:add}
Agents have additive valuations if, an agent $i$, $\forall i \in N$ values any non-empty bundle $A_i$ as $v_i(A_i) = \sum_{k\in A_i} v_{ik} $.
\end{definition}

A valuation instance is said to be \emph{identical} when all agents have the same valuation for all subsets of items, formally, \begin{definition}[\myemph{Identical Valuations}]
 \label{def:identical}
  The valuations are identical if, $\forall i,j \in N, \forall S\subseteq M, v_i(S) = v_j(S)$. 
\end{definition}

Often, agents do not have identical valuations, but they order the items likewise, i.e., agents have the same rank for the items. We call such valuations \emph{Identical Ordering (IDO)}.

\begin{definition}[\myemph{Identical Ordering (IDO)}]
\label{def:ido}
Valuations are IDO when when all agents agree on the same ranking of the items, i.e., for $\forall i \in N$, $v_{i1} \ge v_{i2} \ldots \ge v_{im}$.
\end{definition}
The cardinal cost functions of agents in an IDO instance may still differ. In approval-based settings, agents have binary valuations, where they either approve or disapprove an item.  \begin{definition}[\myemph{Binary Valuations}]
 \label{def:binary}
 The valuations are binary, if $\forall i \in N$, $\forall k \subseteq M$, $v_{ik} \in \{1,0\}$.
 \end{definition}

We next define fairness and efficiency properties considered in this paper.

\subsection{Important Definitions.}

\subsubsection*{\myemph{\textbf{Fairness Related Definitions}}}
The concept of envy-freeness (EF) introduced by \citet{Foley1967ResourceAA} is a well-established notion of fairness. It ensures that no agent envies the bundle of any other agent. Unfortunately, an EF allocation of indivisible items may not exist; for example, when there is a single good and two agents, the agent who doesn't get the good feels envious of the agent that does. The researchers were interested in relaxing the concept of EF in order to limit the envy of every agent. They introduced the idea of $\epsilon$-EF, which ensures that if envy exists, it is at most $\epsilon$. However, $\epsilon$-EF allocations also need not exist. The non-existence of EF and $\epsilon$-EF has prompted researchers to propose additive relaxed notions of EF, such as EF1 and EFX. 

For goods, an allocation is EF1 when all agents value their bundle at least as much as they value another agent's bundle with their \textit{most} valued item removed. EFX is stronger than EF1 and requires that all agents value their bundle no less than the other agents' bundle with their \textit{least} valued item removed. For chores, a similar definition applies, but unlike in goods, a chore is removed from the agent's bundle and then compared with the other agents'. In the case of a combination of goods and chores, EF1 implies that if an agent envies another agent, we can eliminate the envy either by virtually removing the most valued good from the other agent's bundle or removing the least favorite chore from the agent's own bundle. A standard definition is as follows,

\vspace{2.5cm}

\begin{definition}[\myemph{Envy-free (EF) and relaxations}~\cite{aziz2022fairindivisiblegoodsandchores,budish2011EF1,caragiannis2019unreasonable,Foley1967ResourceAA,velez2016fairness}]
For the items (goods or chores), an allocation $A$ that satisfies $\forall i,j \in N$,
\begin{align}
\label{def:ef}
   &  v_i(A_i) \ge v_i(A_j)  \mbox{ is \myemph{EF} } \nonumber\\
    & \begin{rcases}
     v_{ik} < 0,   v_i(A_i \setminus \{ k\}) \ge v_i(A_j);\forall k \in A_i\\ 
v_{ik} > 0,   v_i(A_i ) \ge v_i(A_j \setminus \{ k\});\forall k \in A_j \ 
    \end{rcases}  \mbox{ is \myemph{EFX}}  \nonumber\\
   &  v_i(A_i \setminus \{ k\}) \ge v_i(A_j \setminus \{k\});\exists k \in \{A_i \cup A_j\}  \mbox{ is \myemph{EF1}} \nonumber \\
   &  v_i(A_i) + \epsilon \ge v_i(A_j)  \mbox{ is \myemph{$\epsilon$-EF}} \nonumber \\
   &  v_i(A_i) \ge \alpha \cdot v_i(A_j) \;  \mbox{ is \myemph{$\alpha$-EF}} \nonumber
\end{align}
\end{definition}

In addition to EF and its relaxations, proportional allocations are also well studied in the context of fairness. As introduced by~\citet{propSTEIHAUS}, Proportionality requires that each agent gets at least 1/n of their share of the total value. Furthermore, even proportional allocations may not always exist; for this reason, researchers have considered relaxations. Researchers then looked into $\alpha$-prop, which requires that every agent receives an $\alpha$ fraction of its proportional share. But unfortunately, that solution does not exist as well. Similar to EF, we adapted additive relaxation for proportionality as well. 
Proportionality up to one item, PROP1 requires that every agent is guaranteed to receive their proportionality guarantee if they lose their least valued chore or receive their most valuable good from any other agent's bundle. Proportionality up to any item, PROPX requires that every agent is guaranteed to receive their proportionality guarantee if they lose their most valued chore or receive their least valued good allocated to another agent.
\vspace{1.5cm}

\begin{definition}[\myemph{Proportionality (PROP)} \cite{aziz2020polynomialPROP1PO,baklanov2021propmdontcitebutthenwhatshouldicite,conitzer2017fairpublicdecisionmaking,propSTEIHAUS}]\label{def:prop}
For the items (chores or goods), an allocation $A$ that satisfies $\forall i \in N$,
\begin{align}
&  v_i(A_i) \ge  1/n \cdot v_i(M) \mbox{ is \myemph{PROP}} \nonumber\\
    & \begin{rcases}
v_{ik}>0,      v_i(A_i \cup \{k\}) \ge 1/n \cdot v_i(M);\forall \ k \in \{M \setminus A_i\} \\
v_{ik}<0, v_i(A_i \setminus \{k\}) \ge    1/n \cdot v_i(M);\forall \ k \in A_i
    \end{rcases}  \mbox{ is \myemph{PROPX}}  \nonumber\\ 
 & \begin{rcases}
v_i(A_i \cup \{k\}) \ge 1/n \cdot v_i(M); \exists \ k \in \{M \setminus A_i\}  \mbox{ or,}\\ 
v_i(A_i \setminus \{k\}) \ge  1/n \cdot v_i(M); \exists \ k \in A_i    \nonumber\end{rcases} \mbox{ is \myemph{PROP1} } \nonumber \\
& v_i(A_i \cup \{k\}) \ge  1/n \cdot v_i(M); k \in \max_{j\ne i} \min_{k\in A_j} v_i(A_{jk}) \mbox{ is \myemph{PROPm (only for goods)}} \nonumber \\
 &  v_i(A_i) \ge  \alpha \cdot 1/n \cdot v_i(M) \mbox{ is \myemph{$\alpha$-PROP}} \nonumber
\end{align}

\end{definition}


\smallskip
\noindent{{\myemph{Relationship between EF and PROP.}}}

When agents have sub-additive valuations, Envy-freeness is a stronger notion than proportionality, i.e., EF implies PROP. Similarly, when agents have additive valuations, EF1 implies PROP1. It may seem counter-intuitive at first glance that EFX implies PROPX when agents have additive valuations. Note that EF $\implies$ EFX $\implies$ EF1. PROP $\implies$ PROPX $\implies$ PROPm $\implies$ PROP1. 




Following, we explore Maximin share (MMS) introduced by \citet{budish2011EF1}, extending the concept of Cut and Choose to indivisible goods. Imagine telling an agent to divide the items into $n$ bundles and pick the one they value the least. The agent would divide the items into $n$ partitions so that they maximize the value of the least valued bundle. We call this value their Maximin Share. An allocation satisfies MMS guarantees if every agent receives at least their MMS value.


\begin{definition}[\myemph{Maxmin Share MMS} \cite{budish2011EF1}]
\label{def:mms}
An allocation $A$ is said to be MMS if $\forall i \in N, v_i(A_i) \ge \mu_i$, where
$$\mu_i = \max_{(A_1,A_2,\ldots,A_n) \in \prod_n(M)} \min_{j \in N} v_i(A_j)$$

Introduced by~\citeauthor{procaccia2014mmsdoesntexist}~\cite{procaccia2014mmsdoesntexist}, an allocation $A$ is said to be \myemph{$\alpha$-MMS} if it guarantees $v_i(A_i) \ge \alpha \cdot \mu_i$ when $\mu_i \ge 0$ and $v_i(A_i) \ge \frac{1}{\alpha} \cdot \mu_i $ when $\mu_i < 0$, where $\alpha \in (0,1]$.

\end{definition}


\subsubsection*{\myemph{\textbf{Efficiency Related Definitions}}}

In the previous subsection, we discussed some popular fairness notions. Note that refusing to assign any item to any agent is also Envy-free. However, we also desire efficiency, so fairness is considered in connection with efficiency criteria as well. One of the most frequently studied efficiency criteria in fairness literature is Pareto-Optimality. A Pareto optimal (PO) allocation ensures that there is no other allocation which Pareto dominates, i.e., better for all agents and strictly better for at least one. It is interesting to consider PO and fair allocations.

\begin{definition}[\myemph{Pareto-Optimal (PO)}] 
An allocation $A'$ is said to Pareto dominate allocation $A$, if $\forall i \in N$, $v_i(A'_i) \ge v_i(A_i)$ and $\exists i \in N$, $v_i(A'_i) > v_i(A_i)$. When there is no allocation that Pareto dominates allocation $A$, it is said to be Pareto-optimal.
\end{definition}


\noindent 
We then consider utilitarian welfare, the sum of agents' utilities. On the other hand, Nash welfare corresponds to the product of agents' utilities, and egalitarian welfare, to the minimum of individual agents' utility.
\vspace{1.5cm}
\begin{definition}[] Given an instance $(N, M, \mathcal{V})$, an allocation $A^*$ satisfies,
\begin{align}
    \mbox{\myemph{Maximum}} & \mbox{ \myemph{Utilitarian Welfare}, MUW}(v), \ \mbox{if} \nonumber\\
    &A^* \in \max_{A}\sum_{i=1}^{n} v_i(A_i) \\
    \mbox{\myemph{Maximum}} &\mbox{ \myemph{Nash Welfare}, MNW}(v) \ \mbox{if} \nonumber \\
    & A^* \in \max_{A}\prod_{i=1}^{n} v_i(A_i) \\
    \mbox{\myemph{Maximum}} &\mbox{ \myemph{Egalitarian Welfare}, MEW}(v) \ \mbox{if} \nonumber \\
    & A^* = \max_{A}\min_{i} v_i(A_i)
\end{align}
\end{definition}


We proceed to the next section, summarizing the existential and computational results pertaining to various fairness notions. We discuss various results for goods in Section~\ref{sec:survey_goods} and further categorize them in subsections~\ref{subsec:sg_fair} and~\ref{subsec:sg_fair_eff}, which deal with the results of determining the fair allocation and fair and efficient allocations, respectively. Our discussion of chores is similarly addressed in Section~\ref{sec:chores}, and the combination of goods and chores is described in Section~\ref{sec:combination}.



\section{Indivisible Goods}
\label{sec:survey_goods}


\subsection{Fair Allocation}
\label{subsec:sg_fair}
\subsubsection*{\myemph{\textbf{Envy-Freeness (EF):}}}
As defined in Definition~\ref{def:ef}, EF is a popular and strong notion of fairness. However, EF allocation may not exist for $n\ge2$, for instance, imagine that we have to distribute a single good among $n$ agents. Even when agents have binary additive valuations for goods, the problem of checking whether EF allocation exists or not is \myemph{NP-complete}~\cite{aziz2015fairordinal,hosseini2020fair}. 
Hence, next, we study a prominent relaxation of EF, i.e., EF1 - Envy-free up to one item. The notion of EF1 was introduced by~\citet{budish2011EF1} formally and implicitly it appeared in~\citet{lipton2004approximately}.

\subsubsection*{\myemph{\textbf{Envy-Freeness up to one item (EF1):}}}

As given in Definition~\ref{def:ef}, in the case of goods, Envy-free up to one item (EF1) implies that for any pair of agents $i$ and $j$, if agent $i$ envies agent $j$, the envy can be eliminated by virtually taking out agent $i$'s most valuable item from $j$'s bundle, i.e., $v_i(A_i) \ge v_i(A_j \setminus \{g\}), \; \exists g \in A_j$. EF1 allocation always exists and can be obtained in \myemph{polynomial time}. 

\begin{theorem}[\cite{budish2011EF1,caragiannis2019unreasonable,lipton2004approximately}]
EF1 allocation always exist and can be found in polynomial time.
\end{theorem}

When agents have \myemph{additive valuations}, round-robin algorithm gives EF1 for goods in polynomial time, i.e., $\mathcal{O}(mn\log{}m)$~\cite{caragiannis2019unreasonable}. We formally describe the steps in Algorithm~\ref{algo:roundrobin}.
To allocate items in a round-robin fashion, we select an arbitrary sequence of agents. Each agent selects their favorite items when their turn comes. For any $i,j \in N$, if agent $i$ picks before agent $j$, they will never envy agent $j$. If agent $i$ picks after agent $j$, they may envy agent $j$'s first selected item. Let's assume that we remove agent $j$'s first pick, which was the item it most preferred. Because we removed agent $j$'s first selected item, hypothetical agent $i$ picks first. We know that the agent who comes earlier in the sequence has no envy for the agent who comes after them. Thus, agent $i$ does not envy agent $j$ if we remove agent $j$'s first selected item, i.e., the envy is bounded up to one item.

\begin{tcolorbox}[enhanced,drop fuzzy shadow southwest, colframe=red!20!white!30!black,colback=LimeGreen!10]
\begin{algorithm}[H]
\caption{\myemph{Round Robin}}\label{algo:roundrobin}
\begin{algorithmic}[1]
\State Set $\forall i, A_i = \emptyset$
\State Every agent sort the items in decreasing order
\State Arrange agents in an arbitrary sequence w.l.o.g $\{1, \ldots, n\}$
\While{$M \neq \emptyset$}
\For{$i \leftarrow 1 \textit{ to } n$}
\State ${A_i \leftarrow A_i \cup \mbox{arg}\max_{k \in M} v_{ik}}$
\State $M \leftarrow M \setminus \{k\}$
\EndFor
\EndWhile 
\Return Allocation $A$
\end{algorithmic}
\end{algorithm}
\end{tcolorbox}


When agents have \myemph{general monotone valuations},~\citet{lipton2004approximately} proposed an envy-cycle elimination algorithm that gives EF1 allocation for goods in \myemph{polynomial time} $\mathcal{O}(mn^3)$. This algorithm bounds the envy of any agent by the maximum marginal value of any good; it corresponds to the notion of EF1 for goods. We define the envy graph first. An envy graph of allocation $A$ consists of nodes for each agent and directed edges from agent $i$ to agent $j$ if $i$ envies $j$, i.e., $v_i(A_i) < v_i(A_j)$. The algorithm selects an unenvied agent in each iteration, i.e., an agent with no edges directed towards them, and assigns an arbitrary good. If there are no such agents, there must be cycles of envy, and the agents can exchange bundles until no more cycles remain. Upon receiving the good, other agents may envy this agent; we can eliminate this envy by removing the good they just received since they were previously unenvied. Thus, if agent $i$ envy agent $j$, the envy is bounded up to one good, i.e., by the recently added good in agent $j$'s bundle. After each round of partial allocation, this algorithm ensures that EF1 is satisfied. In contrast,~\citet{berczi2020envy} showed that the envy-cycle elimination algorithm fails to find an EF1 allocation when agents have \myemph{general non-monotone valuations}. However, whether or not EF1 allocation exists for general non-monotone valuations is still an open question.



\begin{tcolorbox}[enhanced,drop fuzzy shadow southwest, colframe=red!20!white!30!black,colback=LimeGreen!10]
\begin{algorithm}[H]
\caption{\myemph{Envy-Cycle Elimination}}\label{algo:cycleenvyelimi}
\begin{algorithmic}[1]
\State Set $\forall i, A_i = \emptyset$
\For{$k \leftarrow 1 \textit{ to } m$}
\State Find an unenvied agent $i$
\State $A_i \gets A_i \cup k$
\If{Envy-Cycle exists}
\State Swap bundles to resolve
\EndIf
\EndFor
\Return Allocation $A$
\end{algorithmic}
\end{algorithm}
\end{tcolorbox}

Apart from EF1, \myemph{weighted EF1 allocation} always exists for goods~\cite{chakraborty2021weighted}, i.e., in presence of asymmetric agents.
Now in order to eliminate envy, EF1 removes the most valuable item. However, the exciting results of EF1 may probably not apply to the real world; i.e., EF1 can be relatively unfair, but the instance may still admit a fairer allocation. 

\begin{tcolorbox}[enhanced,drop fuzzy shadow southwest, colframe=red!20!white!30!black,colback=blue!10]
\begin{example}
\label{ex:divandindiv}
Assume Agastya and Noorie have one Toberlone and 99 Eclairs to divide among themselves. Both prefer Toberlone over Eclairs. The round-robin allocation gives 49 Eclairs, one Toberlone to Noorie, and 50 Eclairs to Agastya. Even though the allocation is EF1, it is unfair to Agastya. 
\end{example}
\end{tcolorbox}

Therefore, it is desirable to strengthen fairness in many cases.~\citet{caragiannis2019unreasonable} proposed the concept of EFX, i.e., Envy-free up to any item as defined in Definition~\ref{def:ef}. Since then, a lot of work has been focused on EFX allocation's existence, but the question is yet to be resolved beyond the case of two agents. The existence of EFX is an open question for $n>3$ for additive valuations and $n>2$ for general valuations. Researchers consider EFX to be the best fairness notion for indivisible items. In contrast to the weaker notion of EF1, the study of EFX existence remains limited and elusive.


\subsubsection*{\myemph{\textbf{Envy-Freeness up to any item (EFX):}}}
As defined in Definition~\ref{def:ef}, in the case of goods, EFX implies that for any pair of agents $i$ and $j$, if $i$ envies $j$, the envy can be eliminated by hypothetically removing any good from $j$'s bundle, i.e., by removing the least valued item for $i$ from $j$'s bundle, i.e., $v_i(A_i) \ge v_i(A_j \setminus \{g\}), \; \forall g \in A_j$. Note that EFX implies EF1. 

\begin{theorem}[\citet{plaut2020EFX}]
When agents have general but identical valuations, a modified Leximin algorithm, i.e., Leximin++, ensures EFX.
\end{theorem}

\citet{plaut2020EFX} showed that when agents have \myemph{general but identical valuations}, a modification of the Leximin, i.e, leximin++ solution is EFX. As mentioned in Preliminaries~\ref{sec:prelims}, in the case of identical valuations, for any pair of agents $i$ and $j$, $\forall S \subseteq M, \; v_i(S) = v_j(S)$. Leximin selects the allocation that maximizes the minimum individual utility; further, if multiple allocations achieve it, it chooses the allocation that maximizes the second minimum individual utility, and so forth. It is well-known that finding the Leximin solution can take \myemph{exponential time}. Plaut modified the Leximin algorithm as follows; Leximin++ selects the allocation that maximizes the minimum individual utility; further, if multiple allocations achieve it, it chooses the allocation that maximizes the size of the minimum utility agent's bundle. Further, if multiple allocations achieve it, it chooses the allocation that maximizes the second minimum individual utility, followed by the cardinality maximum of the second minimum agent's bundle, and so forth.

\begin{tcolorbox}[enhanced,drop fuzzy shadow southwest, colframe=red!20!white!30!black,colback=LimeGreen!10]
\begin{algorithm}[H]
\caption{\myemph{Leximin++ Algorithm}}\label{algo:leximin}
\begin{algorithmic}[1]
\State Set $\forall i, A_i = \emptyset$
\State $A \gets \max_{A' \in \prod_n(M)} \min_{i \in N} v_i(A'_i)$
\While{$|A| > 1$}
\State $i \gets \min_{j \in N} v_j(A_j)$
\State $A \gets \max_{A} |A_i|$
\State $A \gets \max_{A' \in A} \min_{i \in N} v_i(A'_i)$
\EndWhile
\Return Allocation $A$
\end{algorithmic}
\end{algorithm}
\end{tcolorbox}

\begin{remark}
However, Leximin++ may not ensure EFX allocation when agents have distinct valuations
\end{remark}

\begin{theorem}[\citet{plaut2020EFX}]
An EFX allocation exists for two agents with general monotone valuations.
\end{theorem}
Using Leximin++ combined with Cut-and-Choose,~\citet{plaut2020EFX} demonstrated that there exists EFX allocation for two agents. As described in algorithm~\ref{algo:leximinCutandChoose}, the first agent applies Leximin++ to partition the goods into two bundles (i.e., they compute Leximin++ with two copies of themselves). The second agent picks their favorite bundle. As can be seen, the second agent never envies the first agent. Indeed, according to agent 1, the allocation is EFX regardless of what they receive because the partition was created using Leximin++.

\begin{tcolorbox}[enhanced,drop fuzzy shadow southwest, colframe=red!20!white!30!black,colback=LimeGreen!10]
\begin{algorithm}[H]
\caption{\myemph{Leximin++ Combined with Cut-and-Choose}}\label{algo:leximinCutandChoose}
\begin{algorithmic}[1]
\State Run Algorithm~\ref{algo:leximin} with two copies of first agent
\State $A_2 \gets \max{v_2(A_1), v_2(A_2)}$
\State $A_1 \gets M \setminus A_2$
\Return Allocation $A$
\end{algorithmic}
\end{algorithm}
\end{tcolorbox}

Even when two agents have identical submodular valuations,~\citet{plaut2020EFX} showed that finding EFX takes exponential time. On the other hand, the algorithm of~\citet{lipton2004approximately} finds an EF1 allocation in polynomial time for any number of agents with monotone valuations. Thus,  EFX is indeed significantly stronger than EF1.
In addition,~\citet{plaut2020EFX} proposed an algorithm for EFX allocation in polynomial time, i.e., $\mathcal{O}(mn^3)$ when agents have additive valuations with identical ranking, which relies on envy-cycle elimination~\cite{lipton2004approximately}. \citet{plaut2020EFX} showed that this algorithm combined with Cut-and-Choose results in a polynomial-time EFX allocation for two agents with additive valuations. The first agent runs the algorithm with two copies of themselves, and the second agent picks first, just like algorithm~\ref{algo:leximinCutandChoose}. 
\citet{gourves2014near} also presented a polynomial time algorithm for EFX (called as Near Envy-freeness) when for two agents with additive valuations.


A recent study by~\citet{chaudhury2020efx} showed that EFX allocation always exists for instances with \myemph{three agents with additive valuations}. Their rigorous, constructive proof required an extensive and cumbersome case analysis to prove the existence of EFX that involves \myemph{pseudo-polynomial} time complexity. Despite the ongoing efforts, the question of EFX's existence remains unanswered for any valuation system that involves more than three agents.

\begin{theorem}[\citet{chaudhury2020efx}]
EFX always exists in the case of three agents with additive valuations.
\end{theorem}


EFX is arguably the most compelling fairness notion for indivisible items. However, its existence has not been settled and remains one of the major open questions in the fair division. In an attempt to resolve this problem, a range of impressive solutions has been explored, such as considering approximately EFX or EFX with charity, i.e., we leave some low-value items unallocated or solve the cases when agents have special valuation functions. The authors in~\cite{aleksandrov2019mixedmanna,amanatidis2021mnwandEFXstories,babaioff2021fair} studied EFX allocations when agents have restricted valuation functions. \citet{babaioff2021fair} presented a polynomial-time algorithm for finding EFX allocations for \myemph{submodular dichotomous valuations}. \citet{amanatidis2021mnwandEFXstories} proposed a polynomial-time algorithm called \myemph{"MATCH \& FREEZE"} for allocating EFX allocations (they refer to it as EFX$_0$) for 2-value instances, i.e., $\forall i \in N, k \in M, v_{ik} \in I$ and $|I|=2$. In addition, they presented a modified round-robin, in which they reverse the order of agents in the last round, giving EFX when agents have a valuation in an interval, i.e., $\forall i \in N,k \in M, v_{ik} \in [x_i,2x_i]$, where $x_i>0$.



\paragraph*{{\myemph{EFX with charity}}}
We search for partial EFX allocations, which do not require all items to be allocated and satisfy EFX. A trivial allocation involves no items assigned to any agent. Therefore, the goal is to determine allocations with some bound on the set of unallocated items. Authors in~\cite{caragiannis2019envyDonation,berger2021almostEFXwithcharity,chaudhury2021EFXwithcharity} showed that an EFX allocation always exists if we do not allocate at most $n-1$ goods. This result was improved by~\cite{berger2021almostEFXwithcharity} to $n-2$ goods. They also proved that for four agents, EFX could be ensured by leaving at most one good. For general valuations,~\citet{chaudhury2020efx} showed that an EFX allocation \myemph{always exists} with a pool of unallocated items $C$ such that no one envies the pool and $C$ has less than $n$ goods. \citet{caragiannis2019envyDonation} demonstrated that for goods with additive valuations, there exists a partial EFX allocation where all agents get at least half of their MNW allocation value, thus indicating that unallocated goods are low-valued.
In~\cite{caragiannis2019envyDonation}, the authors conjectured that, 
"In particular, we suspect that adding an item to an allocation problem (that provably has an EFX allocation) yields another problem that also has an EFX allocation with at least as high Nash welfare as the initial one." Nevertheless,~\citet{chaudhury2020efx} disproved the conjecture by displaying an example of a partial EFX allocation with a higher Nash welfare than a complete EFX allocation. In parallel to this line of research, many impressive results show the existence of approximate EFX allocations.

\paragraph*{{\myemph{Approximately EFX}}}
\citet{plaut2020EFX} proposed approximate EFX, which implies that for any pair of agents $i$ and $j$, $v_i(A_i) \ge \alpha \cdot v_i(A_j \setminus \{g\}), \; \forall g \in A_j$, where $\alpha>0$. Note that $\alpha$-EFX and EF1 are not comparable, meaning none of the properties implies the other. \citet{plaut2020EFX} showed that 1/2-EFX allocations always exist when players have sub-additive valuations using a modified envy-cycle elimination method. \citet{amanatidis2020multiplebirds} proved the existence of \myemph{$0.618$-EFX} allocations for additive valuations. For $\epsilon \in (0, 1/2]$,~\citet{chaudhury2021improving} showed that there is always an $(1 - \epsilon)$-EFX allocation with a sub-linear number of unallocated goods and a high Nash welfare. \citet{amanatidis2021mnwandEFXstories} also proposed $\alpha$-EFX-value notion.

The envy-freeness concept doesn't always make sense, for instance, in the context of \myemph{Public Decision Making}.
As explained in the Introduction, Agastya and Noorie are supposed to purchase a new refrigerator; Agastya likes Samsung, and Noorie likes LG. Regardless of what kind of refrigerator they buy, the concept of envy is no longer applicable. In the next section, we discuss the results related to proportionality for indivisible goods.
We now move to another intuitive fairness notion introduced by~\citet{propSTEIHAUS}, \myemph{proportionality}. 


 \begin{figure}[H]
     \centering
    \includegraphics[width=\textwidth]{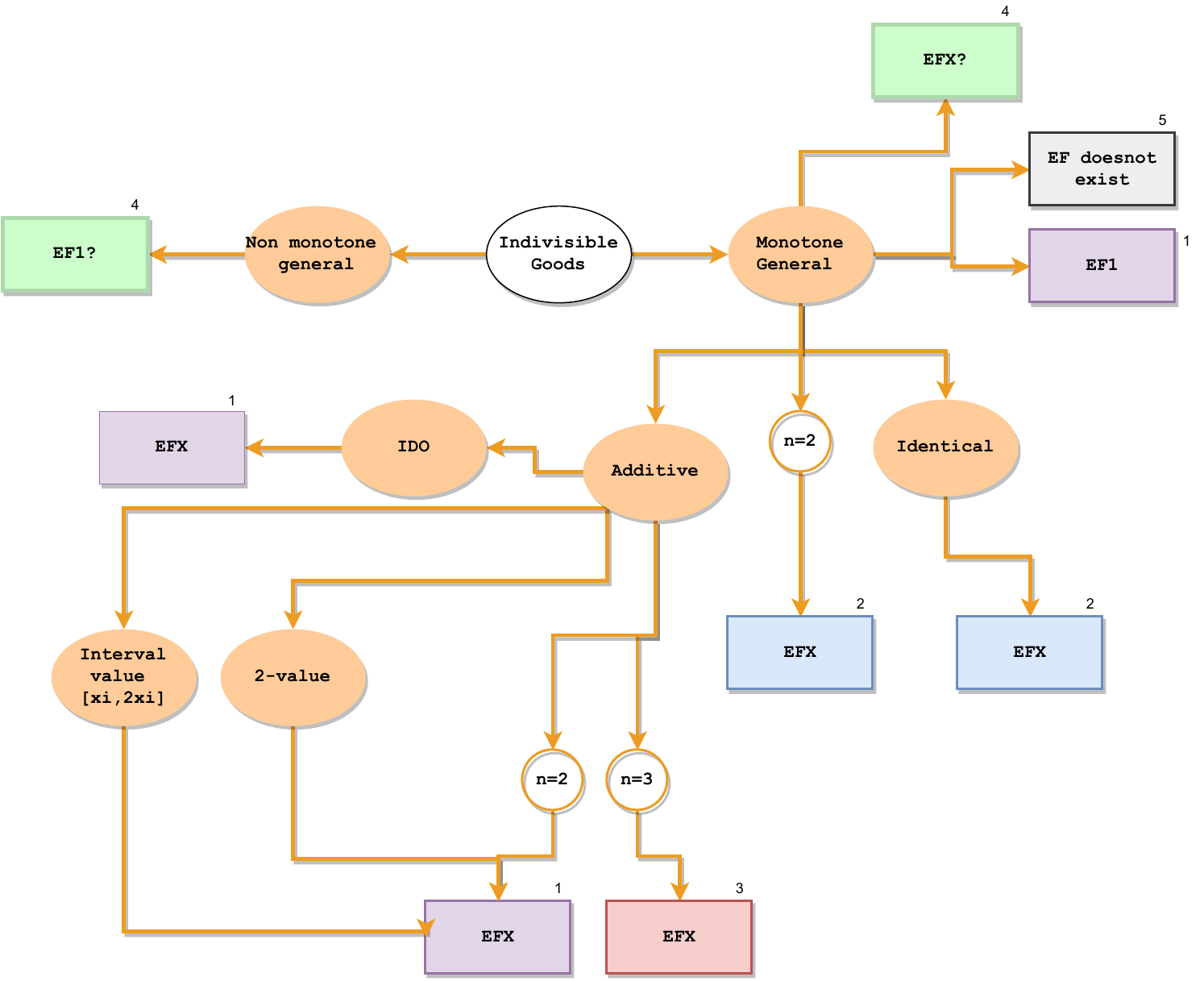}
     \caption{Summary of Envy-freeness fairness notion for indivisible goods. The Violet box (number 1) represents the polynomial-time algorithm. The blue box (number 2) represents exponential time. The red box (number 3) represents Pseudo-polynomial time. The green box (number 4) represents an open question. The grey box (number 5) represents such an allocation does not exist.}
     \label{fig:EFindivisiblegoodssummary}
 \end{figure}


\subsubsection*{\myemph{\textbf{Proportionality (PROP):}}}
To attain proportionality, each agent would have to receive at least $1/n$ of their value for all items, as stated in Definition~\ref{def:prop}. However, it is known that it is not always possible to assign indivisible items proportionality. The allocation of a single good among agents cannot ever be proportional. Note that EF allocation is always PROP with sub-additive valuation. For $\alpha>0$, $\alpha$-proportional allocation may not always exist; for example, in the distribution of single good among two agents, one receives a utility of zero. Naturally, researchers explored the \myemph{additive approximations of PROP}, i.e., PROP1, PROPm, and PROPX. In any situation, we look for the fairest allocation possible, and if it is not possible, we relax our criteria in a way that maintains fairness.

\subsubsection*{\myemph{\textbf{Proportionality up to one item (PROP1):}}}
\citet{conitzer2017fairpublicdecisionmaking} introduced the notion of PROP1 in the setting of Public Decision Making, which is more generic than indivisible item allocations. Along the lines of EF1, PROP1 requires each agent to receive a utility at least their proportional share if we add the largest good allocated to another agent to their bundle, as stated in Definition~\ref{def:prop}. Each agent receives its proportional share after hypothetically including one extra good from another agent's allocation into its bundle. 

Under additive valuation, an EF1 allocation is also a PROP1 allocation; therefore, every algorithm that gives EF1 also gives PROP1. In the setting of Public Decision Making,~\citet{conitzer2017fairpublicdecisionmaking} shows that a PROP1 allocation always exists in \myemph{polynomial time}. In addition to fairness,~\citet{aziz2022fairindivisiblegoodsandchores} demonstrated that one could achieve another requirement besides fairness: connectivity. They modeled the situation in which items are placed on a path, and each agent wants a connected bundle of the path. In many scenarios, finding a connected set of items is essential. For example, the items could be a set of rooms along a corridor, and the agents could be research groups seeking adjacent rooms. The authors showed that contiguous proportional allocations of goods exist for additive utilities and that they can be computed in polynomial time. In the later sections, we will discuss how the combination of efficiency and fairness becomes more intriguing in these contexts.

\begin{theorem}[\citet{aziz2022fairindivisiblegoodsandchores}]
Connected PROP1 exists in polynomial time for indivisible goods.
\end{theorem}


\subsubsection*{\myemph{Proportionality up to any item (PROPX):}}

PROPX requires each agent to receive a utility at least their proportional share if we add the small good allocated among the remaining agents to their bundle, as stated in Definition~\ref{def:prop}. Each agent receives its proportional share after hypothetically including one extra smallest good from the remaining agents' allocations into its bundle. The PROPX notion is stronger than PROP1. It is also important to note that EFX is weaker than PROPX. While EFX exists for three agents with additive valuations~\cite{chaudhury2020efx},~\citet{aziz2020polynomialPROP1PO} presented the following counter-example where PROPX \myemph{doesn't exist} with three agents having identical valuations over five goods, i.e., they all value each good 3,3,3,3, and 1 respectively. 

In contrast to PROP1, which is easy to satisfy and much weaker, PROPX is much more demanding and is not known to exist even for small instances. Therefore this led researchers to explore alternate ways to relax proportionality. 

\subsubsection*{\myemph{Proportionality up to the maximin item (PROPm): }}

The problem is that PROP1 is too weak, and PROPX may not exist; there has to be a fairness concept that is stricter than PROP1 but weaker than PROPX;~\citet{baklanov2021propmdontcitebutthenwhatshouldicite} introduced the notion of PROPm. PROPX requires each agent to receive a utility worth at least their proportional share if we add the largest good among the set of smallest goods allotted to the remaining agents to their bundle, as stated in Definition 1. In other words, PROPm considers the least valued item (from i’s perspective) in each of the other agent’s bundles and then takes the highest value among them. Note that PROPX implies PROPm implies PROP1.

\citet{baklanov2021propmdontcitebutthenwhatshouldicite} presented that PROPm exists at least for up to five agents with additive valuations, though the algorithm is exponential in terms of the number of items. There is a limit of five agents, not because of the rigid limitations of the constructive proof of their algorithm, but because as agents increases, the complexity of analyzing all cases increases. As a result, they conjectured that PROPm will always exist for any number of agents or items. Additionally,~\citet{baklanov2021propmdontcitebutthenwhatshouldicite} explored various relaxations of PROP, such as allowing the value to be added to be equal to a median, mode, or minimax value of the agent. However, their results showed that this fails to exist for as little as three agents. They presented a counter-example with three agents having identical additive valuation over seven items, i.e., all value the items $1-6\epsilon,\epsilon,\epsilon,\epsilon,\epsilon,\epsilon$ and $\epsilon$, respectively. It is easy to verify that there always exists an agent with utility at most 3. Hence, it violates approximate proportionality with the mean ($<0.25$), the median ($\epsilon$), the mode ($\epsilon$), and the minimax item value ($\epsilon$). Later on,~\citet{baklanov2021propmfinalllyy} demonstrated that PROPm allocations exist in \myemph{polynomial time}. 

\begin{theorem}[\citet{baklanov2021propmfinalllyy}]
PROPm exists, and we can find it in polynomial time in terms of the number of agents and goods. 
\end{theorem}

The heart of the algorithm lies in breaking the allocation problem into subparts and finding PROPm in each of them, i.e., it recursively solves PROPm by splitting into agent- and item-disjoint sub-problems. This allocation is also PROPm for the original instance. The algorithm starts by selecting an arbitrary divider agent $i$, and this agent sorts the items in non-decreasing order and partitions $m$ items into $n$ bundles, $(S_1,S_2,\ldots,S_n)$, such that $\forall j \in [n], v_i(S_j) = 1/(n-j+1) \cdot v_i(M \setminus (\cup_{i=1}^{j-1}S_i))$. The algorithm then evaluates how other agents value these bundles, splits the problem into two, and assigns a bundle to this divider agent. Initially, this is one problem; hence, the proportional Decomposition set $D$ is empty. We set $N_d$ as an empty set representing the agents interested in $D$, i.e., their valuation of the items in $D$ is proportional. We set $N_r$ as $N^{-i}$ representing the remaining agents. The algorithm iterates for $n$ rounds; in any round $t$, it computes $c$ as the number of agents in $N_r$ who value the first $t$ bundles $> t/n$. If $c$ is $0$ and $|N_d|$ is less than $t$, it assigns $S_t$ to the divider agent. It further decomposes the agents-items into two sub-problems involving $N_d$ agents composed of $t-1$ agents with $D$ items consisting of the first $t-1$ bundles and $N_r$ agents composed of $n-t$ agents with the last $n-t$ bundles. If $c>0$, there is an over-demand of the initial set of bundles, so the algorithm calls a subroutine UpdateDecomposition. The subroutine uses graph analysis (novel graph representation) to update $D$ by either reducing $c$ by one or increasing the agents in $N_d$ by one.


 \begin{figure}[h]
     \centering
    \includegraphics[width=0.5\textwidth]{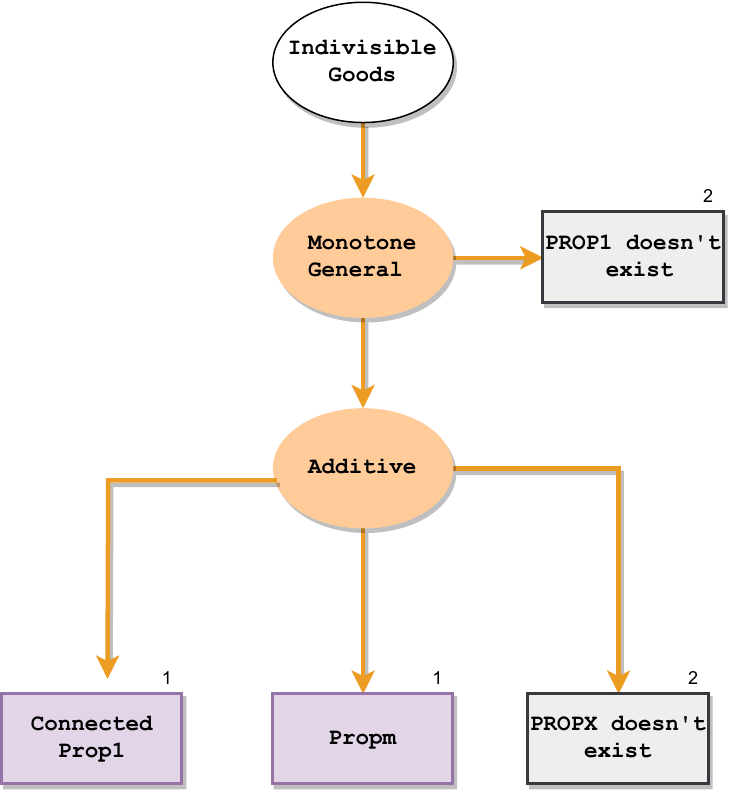}
     \caption{Summary of Proportionality fairness notion for indivisible goods. The Violet box (number 1) represents the polynomial-time algorithm. The grey box (number 2) represents such an allocation does not exist.}
     \label{fig:Propindivisiblegoodssummary}
 \end{figure}

Additionally, an alternate relaxation that has received considerable attention is~\citet{budish2011EF1}'s maximin share (MMS). We study MMS allocation in detail in the following section.

\subsubsection*{\myemph{\textbf{Maximin Share (MMS):}}}
In essence, MMS is an extension of the well-known \myemph{``Cut-and-Choose''} algorithm used in solving the cake-cutting problem. Let's say we ask an agent to divide $m$ items into $n$ bundles and take the bundle that they're least interested in. This risk-averse agent would divide the bundles to maximize the minimum utility, which is the MMS share of the agent. An MMS allocation guarantees every agent their MMS share. Note that MMS is a weaker fairness property than proportionality; a proportional allocation is always MMS. Even though MMS is weaker, MMS is still too demanding in the case of indivisible items.

Despite its appealing formulation, MMS does have a computational disadvantage. For one, even the computation of the maximin share for an agent with additive valuations is an \myemph{NP-Complete} problem~\cite{bouveret2016characterizingconflictsinFD}. 
And computing an MMS allocation is strongly \myemph{NP-Hard}. The problem is weakly NP-hard even for two agents~\cite{azizmmschores,bouveret2016characterizingconflictsinFD}. However, a PTAS for computing MMS exists~\cite{woeginger1997ptas}. Originally,~\citet{woeginger1997ptas} gave PTAS to compute a maximum partition for a particular agent within the context of job scheduling. However, the problem is identical to computing a maximin partition for the given agent. 

\citet{bouveret2016characterizingconflictsinFD} addressed the question of whether MMS allocation exists for indivisible items. While they answered the question for special cases, they left the question of MMS allocation for additive valuations open. 
Their results show that when agents have additive valuations, MMS allocations exist in the following cases: (i) \myemph{Binary valuations}, (ii)  \myemph{For two agents}, and (iii) \myemph{$m \le n+3$}. 
When agents have binary valuations,~\citet{bouveret2016characterizingconflictsinFD} showed that an MMS allocation exists in polynomial time and can be found by employing a round-robin algorithm with an alternating sequence, whereby each agent iteratively selects their most favorite item from the remaining ones. By using polynomial-time case analysis,~\citet{bouveret2016characterizingconflictsinFD} demonstrated that there exists MMS allocation for instances having $n$ agents and $n+3$ goods.
An MMS allocation exists in instances with two agents with additive valuations using the ``Cut-and-Choose'' protocol.~\citet{bouveret2016characterizingconflictsinFD} showed that it is challenging to find an MMS allocation when agents have similar preferences, i.e., the more conflict, the more difficult it is to find an MMS allocation. The authors proved that if one can find an MMS allocation for an identical preference instance, every permutation derived from that preference instance can also be allocated as an MMS allocation.

\citet{bouveret2016characterizingconflictsinFD} showed that an MMS allocation \myemph{need not exist} when agents have general valuations, even in the case of two agents, and left the existence of MMS allocation for additive valuations as an open question. 
In the case of additive valuations, despite the experimental results of~\cite{bouveret2016characterizingconflictsinFD}, which had MMS allocations consistently,~\citet{procaccia2014mmsdoesntexist} presented an intricate example in which every allocation fails to achieve MMS guarantees.
Thus, MMS is not guaranteed to exist in instances with more than two agents with additive valuations. The authors of~\cite{feige2021tightnegativeexampleformms,kurokawa2016canmmsbeguaranteed,procaccia2014mmsdoesntexist} provided counter-examples to demonstrate that the MMS allocation for goods may not always exist, even with additive valuations.~\citet{kurokawa2016canmmsbeguaranteed} provided an example involving $m$ that is linear in $n$, while~\citet{procaccia2014mmsdoesntexist} provided an example involving $m$ that is exponential in $n$.

\begin{theorem}
Given a resource allocation problem $(N,M,\mathcal{V})$, there exists $M$ and (additive) valuations such that no allocation is MMS for any $n\ge3$, 
\end{theorem}

Because MMS need not exist, naturally, we seek relaxation in the allocation. The goal is to construct an efficient algorithm to provide approximate MMS guarantees. Fortunately, a \myemph{multiplicative approximation} of MMS always exists. We discuss $\alpha$-MMS in the following section.

\vspace{1cm}


\subsubsection*{\myemph{\textbf{$\alpha$-Maximin Share (MMS):}}}

\citet{procaccia2014mmsdoesntexist} introduced a new concept of approximate maximin share ($\alpha$-MMS) to solve this problem. An $\alpha$-MMS allocation guarantees every agent at least the $\alpha$ fraction of the MMS share. Note that all the results mentioned below are in the presence of additive valuations for MMS allocation.


The papers~\cite{Amanatidismms2017,barmanapproxmms2020,procaccia2014mmsdoesntexist} showed that 2/3-MMS for goods always exists. Paper~\cite{gargmms2019,ghodsi2018mms} showed that 3/4-MMS for goods always exists. Authors in~\cite{gargmms2019} provide an algorithm that guarantees \myemph{$3/4+1/12n$}-MMS for goods.

\citet{procaccia2014mmsdoesntexist} proved that the 2/3-MMS allocation for goods always exists and devised an exponential-time algorithm. The algorithm enables one of the agents to divide the items as its maximum partition and creates a bipartite graph (between agents and bundles so that edges exist if a bundle is acceptable to the agent). Using Hall's Theorem, they showed that such a graph has a perfect matching, and the algorithm runs recursively on the remaining agents and items. If $n$ is constant, the (PTAS) algorithm runs in polynomial time. They also showed that in the case of three and four agents, 3/4-MMS allocation always exists using the same algorithm.

\citet{Amanatidismms2017} presented a PTAS algorithm that computes $2/3-\epsilon$-MMS allocation, polynomial for any number of agents and goods. They redesigned the algorithm presented in~\cite{procaccia2014mmsdoesntexist} by manipulating matchings of the bipartite graph. Further, they proved that in the case of three agents, there always exists 7/8-MMS allocation and $7/8-\epsilon$ in polynomial time using the PTAS algorithm. Lastly, when the valuation of every agent for each item, i.e., $\forall i,k, v_{ik} \in \{0,1,2\}$, there always exists MMS allocation and presented a polynomial-time algorithm, a variant of a round-robin algorithm. They also provided a much simpler and faster algorithm that computes at least 1/2-MMS, based on a round-robin algorithm with a slight modification of allocating the most valuable goods first.~\citet{barmanapproxmms2020} developed an algorithm that achieves the same level of approximation guarantee as these results. \citet{gargmms2019} presented a simple polynomial-time algorithm for 2/3-MMS allocation with straightforward analysis.

Later,~\citet{ghodsi2018mms} proved that 3/4-MMS always exists and gave a PTAS algorithm for allocating $3/4-\epsilon$-MMS in polynomial time. This result is exciting because~\citet{procaccia2014mmsdoesntexist} proved that a 2/3-MMS allocation was tight, which means using their algorithm would not guarantee better performance. Furthermore,~\citet{ghodsi2018mms} demonstrated that the 4/5-MMS allocation always exists if there are four agents as well as $4/5-\epsilon$ in polynomial time using the PTAS algorithm. Then~\citet{garg2021improved} presented a polynomial-time algorithm for finding the 3/4-MMS allocation of goods. The most recent results of~\citet{garg2021improved}'s work indicates that $3/4+1/12n$ MMS always exists.

 \begin{figure}[H]
     \centering
    \includegraphics[width=0.9\textwidth]{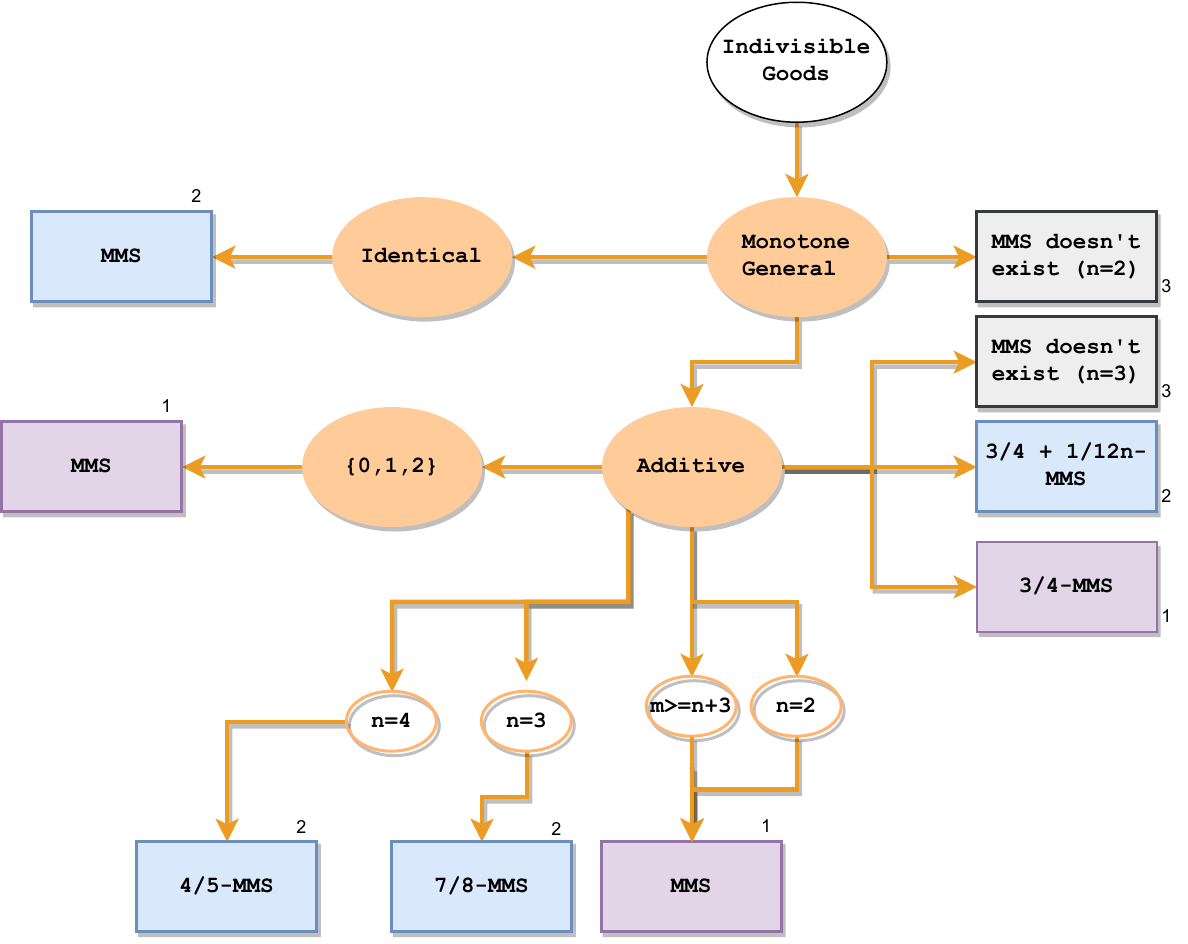}
     \caption{Summary of MMS fairness notion for indivisible goods. The Violet box (number 1) represents the polynomial-time algorithm. The blue box (number 2) represents exponential time. The grey box (number 3) represents such an allocation does not exist.}
     \label{fig:EFindivisiblegoodssummary}
 \end{figure}

We describe the latest algorithm in detail, i.e., the polynomial-time algorithm for \myemph{3/4-MMS} allocation. Except for that, we simply skim over other algorithms briefly. We now talk about two sub-routines.

\smallskip
\noindent{{\myemph{Normalized Valuations}}}
$\alpha$-MMS is scale invariant, i.e., if $A$ is $\alpha$-MMS allocation for the valuation profile $\mathcal{V}$, and if we consider a scaled valuation profile $\mathcal{V}'$, i.e., $\forall i \in N,k \in M, v'_{ik}=c \cdot v_{ik}$, $A$ is $\alpha$-MMS for $\mathcal{V}'$. For simplicity, we scale the valuations such that $\forall i, \mu_i \le 1$, i.e., $v_i(M)=|N|$.

\smallskip
\noindent{{\myemph{Bag Filling.}}}
\citet{ghodsi2018mms} introduced bag filling to allocate low-value goods as an extension of the Moving Knife Procedure (allocates divisible goods proportionally). The algorithm begins by filling an empty bag with arbitrary goods until it satisfies some agent, that is until the agent calls for the bag; after that, the algorithm assigns the bundle to the agent and starts refilling an empty bag again, and repeats the process with the remaining agents. When agents value each item less than $\delta \mu_i$, i.e., $v_{ik} \le \delta \mu_i$, the bag filling algorithm gives $(1-\delta) \mu_i$-MMS allocation. 
The intention is that after an agent receives a bundle, enough goods are left for the rest of the agents to meet their $(1-\delta) \mu_i$-MMS allocations. When assigning a bundle to an agent, the other agents value it less than $(1-\delta) \mu_i$ (since they didn't call for the bundle), so the valuation of the rest of the agents for remaining goods is at least $\frac{n-1}{n}\cdot v_i(M)$, which is proportionally more than before assigning the bundle. We can say that the agents who pick bundles early leave more goods than they take. For normalized valuations (i.e. $\forall i \in N$, $v_i(M) = n$), if $\forall i$ and $\forall k$, $v_{ik} \le \delta$, the bag filling guarantees $(1-\delta)$-MMS allocation, where $0 \le \delta \le 1/2$. 

\begin{tcolorbox}[enhanced,drop fuzzy shadow southwest, colframe=red!20!white!30!black,colback=LimeGreen!10]
\begin{algorithm}[H]
\caption{\myemph{Bag Filling Algorithm}}\label{algo:roundrobin}
\begin{algorithmic}[1]
\State Normalize valuations
\State Set $\forall i, A_i = \emptyset$, $B=\emptyset$
\For{$k \leftarrow 1 \textit{ to } m$}
\If{$\exists i \in N,v_i(B \cup k) > (1-\delta)$}
\State $A_i \gets B$
\State $B \gets \emptyset$
\EndIf
\State $B \gets B \cup k$
\EndFor
\Return Allocation $A$
\end{algorithmic}
\end{algorithm}
\end{tcolorbox}

\smallskip
\noindent{{\myemph{Valid Reduction.}}}
If we remove an agent $i$ and an item $k$ from an instance $\mathcal{I}=\langle M,N,\mathcal{V} \rangle$, it doesn't reduce MMS guarantees for the remaining agents. We mean to say that in $\mathcal{I}'=\langle M \setminus \{i\}, N \setminus \{k\}, \mathcal{V} \rangle$, $\mu'_i \ge \mu_i$, i.e., the MMS value in the new reduced instance is at least the MMS value in the original instance. For high valued items, i.e., $\exists i,k, v_{ik}\le\alpha\mu_i$, we assign item $k$ to agent $i$. After assigning this high-valued item, we still have enough goods, i.e., the MMS value of each agent in the reduced instance is at least the MMS value in the original instance. Our goal is to assign alpha-MMS, i.e., the idea is to assign the high-value items firstly, i.e., giving an agent an item if they value more than alpha mu. We reach the stage where the instance is irreducible, i.e., none of the remaining agents have any remaining high-value items. We say an instance is $\alpha$-irreducible, if there $\not\exists i \in N$, $k\not\exists M$, such that $v_{ik}\ge \alpha\mu_i$.

The core concept lies in using these two algorithms to ensure $\alpha$-MMS, i.e., firstly assign high-valued items using valid reduction and further assign low-valued and medium-valued items using a modified bag filling.

\smallskip
\noindent{{\myemph{A polynomial-time 1/2-MMS}}}
We start by normalizing the valuations for all the agents. We assign an item to an agent if they value it more than 1/2. We break ties arbitrarily. Note that when we assign an item to an agent, in the reduced instance, the new MMS value of all the remaining agents is at least as much as in the original instance. Once the instance becomes 1/2-irreducible, i.e., if there are remaining agents, they all value the remaining items less than 1/2, we assign the rest of the goods using the bag filling algorithm.  

\smallskip
\noindent{{\myemph{Ordered Instances}}}
Further, there's another core concept used by these algorithms, which was first noticed by~\citet{bouveret2016characterizingconflictsinFD}. They showed that it is challenging to find an MMS allocation when agents have similar preferences, i.e., the more conflict, the more difficult it is to find an MMS allocation. The authors proved that if one can find an MMS allocation for an identical preference instance, i.e., an ordered instance, every permutation derived from that preference instance can also be allocated as an MMS allocation. Formally, we say an instance is an ordered instance \emph{iff} there exists a total ordering over the set of goods, i.e., $\forall i \in N, v_{i1} \ge v_{i2} \ge \ldots \ge v_{im}$.
Further,~\citet{bouveret2016characterizingconflictsinFD,barmanapproxmms2020} provided a reduction from any arbitrary instance $I = (N, M, \mathcal{V})$ to an ordered instance $I'$, and showed that if $A'$ satisfies $\alpha$-MMS guarantees for $I'$, we can find allocation $A$ for $I$ in polynomial time derived from $A'$ in polynomial-time. 

Given an instance $I = (N, M, \mathcal{V})$, we construct an ordered instance $I' = (N, M, \mathcal{V}')$ as follows,
\begin{tcolorbox}[enhanced,drop fuzzy shadow southwest, colframe=red!20!white!30!black,colback=LimeGreen!10]
\begin{algorithm}[H]
\caption{\myemph{Converting into Ordered Instance}}\label{algo:convertintoordered}
\begin{algorithmic}[1]
\For{$k \leftarrow 1 \textit{ to } m$}
\For{$i \leftarrow 1 \textit{ to } n$}
\State  $k^* \gets \textit{agent } i\textit{'s } \textit{$k^{th}$ most valuable good}$
\State $v'_{ik} \gets v_{ik^*}$
\EndFor
\EndFor
\Return Valuation $\mathcal{V}'$
\end{algorithmic}
\end{algorithm}
\end{tcolorbox}
Now given an allocation $A'$ which satisfies $\alpha$-MMS for $I'$, we construct $A$ which satisfies the same for $I$ as follows,
\begin{tcolorbox}[enhanced,drop fuzzy shadow southwest, colframe=red!20!white!30!black,colback=LimeGreen!10]
\begin{algorithm}[H]
\caption{\myemph{$\alpha$-MMS for Original Instance $I$}}\label{algo:alphammsfororginialinstace}
\begin{algorithmic}[1]
\State $A=(\emptyset,\emptyset,\ldots,\emptyset)$, and $R \gets M$
\For{$k \leftarrow 1 \textit{ to } m$}
\State Pick the agent $i$ who has item $k$ in $A'$
\State $g \gets \textit{argmax}_{k \in R} v_{ik}$
\State $A_i \gets A_i \cup \{g\}$ and $R \gets R \setminus \{g\}$
\EndFor
\Return Valuation $\mathcal{V}'$
\end{algorithmic}
\end{algorithm}
\end{tcolorbox}
It is easy to verify that this whole procedure is a polynomial-time algorithm. From now, we only need to address the setting of ordered instances, i.e., identical preferences, and using a polynomial time algorithm, we can compute $\alpha$-MMS for any arbitrary additive instances.
Observe that the MMS values of an agent in $I$ and $I'$ are the same because it neither depends on the order of the items nor on other agents’ valuations. 

\smallskip
\noindent{{\myemph{3/4 MMS when we know the $\mu$ values}}}

We now discuss the algorithm for computing 3/4-MMS allocation. We pre-process our instances such that they are normalized and ordered. We assume that we know the MMS value of each agent, and we scale the valuation such that the MMS value becomes 1 for all. Note that this is a crucial assumption; later, for the strongly polynomial time algorithm for exact 3/4, we will change the way we normalize the valuations.

For ease of exposition, we abuse notation and use $M$ and $N$ to denote the set of unallocated items and the set of agents who have not yet received any bundle.
The basic structure of the algorithm closely resembles that of valid reduction and bag filling.

The algorithm first assigns high-valued items. In the algorithms, they make it a simple process of greedy assignment by leveraging the pigeonhole principle to make valid reductions. They define bundles $S_1:=\{1\}$, $S_2:=\{n,n+1\} (\emptyset, \textit{if} m \le n)$, $S_3:=\{2n-1,2n,2n+1\} (\emptyset, \textit{if} m \le 2n)$, and $S_4:=\{1,2n+1\}$. $S_1$ has the highest valued item in $M$, $S_2$ has the $n^{th}$ and $(n+1)^{th}$ highest valued items in $M$, and so on. If any of these bundles value $3/4$ to an agent, then it is a valid reduction. We assign the lowest index bundle in $\{S_1,S_2,S_3,S_4\}$ to agent $i$ such that it is valued at least 3/4.  Note that after each valid reduction, we update $M$ and $N$, and these bundles will change accordingly. 

Once the instance becomes $\alpha$-irreducible, i.e., for the remaining agents $i \in N$, $S \in \{S_1,S_2,S_3,S_4\}$, and $\forall i, S, v_i(S) < 3/4$, we proceed to the modified bag filling algorithm for the rest of the agents. Let $J_1:=\{1,2,\ldots,n\}$ and $J_2:=\{n+1,\ldots,2n\}$. Next, we initialize $n$ bags as follows: $B = \{B_1,B_2,\ldots,B_n\}$; where $B_k = \{k,2n-k+1 \}, \forall k$. Each bag contains one item from $J_1$ and one item from $J_2$ such that from $B_1$ to $B_n$ value of items from $J_1$ decreases and the value of items from $J_2$ increases. Now in each round $k$, it starts a new bundle $T$ with $T \gets B_k$. If there is
an agent who values $T$ to be at least 3/4, then assign $T$ to such an agent. Otherwise, keep adding goods from $M \setminus \{J_1 \cup J_2\}$ to $T$ one by one until an agent with no bundle assigned to her values $T$ at least
3/4. The algorithm allocates $T$ to that agent, and if there are multiple such agents, it chooses one
arbitrarily. The complete details are given in the Algorithm~\ref{algo:34MMS}. 

However, note that this algorithm requires a detail of the exact MMS value of all the agents, which is NP-Hard to compute. The authors further modified the algorithm in such a that it is strongly polynomial in time and computes exact 3/4-MMS allocation with the same underlying concepts.
The key modification is the way we normalize valuation, allocation of $S_4$ bundle, and use the average as an upper bound for the MMS value.

\begin{tcolorbox}[enhanced,drop fuzzy shadow southwest, colframe=red!20!white!30!black,colback=LimeGreen!10]
\begin{algorithm}[H]
\caption{\myemph{Exact 3/4-MMS Algorithm}}\label{algo:34MMS}
\begin{algorithmic}[1]
\State Assume MMS values are known, alpha, 
\State Normalize the Valuations such that $\forall i, \mu_i=1$
\State Using Modified Valid Reduction now
\State $S_1 \gets \{1\}$, $S_2 \gets \{n,n+1\}, S_3\gets\{2n-1,2n,2n+1\}, S_4\gets\{1,2n+1\}$
\State $\forall S \subseteq M$, we define $\tau(S) := \{i \in N : v_i(S) \ge \alpha\}$
\While{$(\tau(S_1)\cup\tau(S_2)\cup\tau(S_3)\cup\tau(S_4))) \ne \emptyset$}
\State $S \gets$ lowest index in $\{S_1,S_2,S_3,S_4\}$ which satisfies $\alpha$
\State agent $i$ in $\tau(S)$
\State $A_i \gets S$
\State $M \gets M \setminus S$; $N \gets N \setminus \{i\}$
\EndWhile
\State Initialize bags $B$, where $B_k = \{k,2n-k+1 \}, \forall k$
\State $R \gets M \setminus J$
\For{$k \leftarrow 1 \textit{ to } n$}
\State $T \gets B_k$
\While{$v_i(T)<\alpha$}
\State $g \in R$
\State $T \gets T \cup \{j\} ; R \gets R \setminus \{j\}$
\EndWhile
\State $\exists i$, $v_i(T)\ge\alpha$, $A_i \gets T$, $N \gets N \setminus i$
\EndFor
\Return Valuation $\mathcal{V}'$
\end{algorithmic}
\end{algorithm}
\end{tcolorbox}
\citet{garg2021improved} proved that existence of $\frac{3}{4} + \frac{1}{12n}$ using Algorithm~\ref{algo:34MMS}. They showed that this bound is tight. They showed that extending this approach for improved $\alpha$ would be challenging.

Apart from additive valuations for indivisible goods, MMS allocations have been studied extensively in various other settings.~\citet{farhadi2019fairwithasymmetricagents} studied MMS allocation for asymmetric agents.~\citet{seddighin2019externalities} explored MMS allocation in the presence of externalities.~\citet{amanatidis2017truthful,amanatidis2016truthful,aziz2019strategyproof,barman2019fair} studied MMS allocation for strategic agents.~\citet{barmanapproxmms2020,ghodsi2018mms,li2021fair,gourves2019maximin,biswas2018fair} studied MMS allocation beyond additive valuations.

Next, we discuss fair and efficient algorithms for indivisible goods.

\subsection{Fair and Efficient Allocation}
\label{subsec:sg_fair_eff}
For various reasons, we strive to seek fair, effective allocations. We want to ensure fairness, but we also want to ensure the items go to the most deserving; however, we encounter various difficulties when doing so.
Within the PROP1 allocation set, finding an allocation that is utilitarian-maximum is NP-hard~\cite{aziz2019constrained}. The problem of making Pareto improvements to PROP1 allocation is computationally challenging~\cite{aziz2016optimal,keijzer2009complexity}. It is possible that a Pareto improvement over a PROP1 allocation will not even satisfy PROP1. In this section, we summarize the literature on efficient fair allocation for indivisible goods. 

\subsubsection*{\myemph{Efficient EF1 allocations}}
Our aim in this section is to discuss the results regarding efficient allocation of EF1.

\paragraph*{\myemph{EF1 + PO}}
Recall that an allocation is said to be Pareto optimal (PO) if it is not Pareto dominated by any other allocation, i.e., there is no other allocation in which every agent is better off and at least one agent is strictly better off. Under additive valuations,~\citeauthor{caragiannis2019unreasonable}~\cite{caragiannis2019unreasonable} has shown that EF1 and PO allocations always exist. As shown in their study, the MNW allocation (i.e., an allocation that maximizes Nash social welfare) satisfies both EF1 and PO. This is challenging because the complexity of computing an allocation like this is high, i.e., even though finding EF1 is polynomial time, even for additive valuations, polynomial-time algorithms are not available to find EF1 and PO. Computing MNW allocation is APX-hard~\cite{lee2017apxcomplexity}. In addition,\citeauthor{barman2018findingEF1PO}~\cite{barman2018findingEF1PO} proposed a \myemph{pseudo-polynomial time} algorithm for determining PO and EF1 allocations.

\paragraph*{\myemph{EF1 + USW}}
Among a set of utilitarian maximal allocations,~\citeauthor{azizEF1PROP1plusUSW}~\cite{azizEF1PROP1plusUSW} proved that it is strongly NP-hard to find an allocation that is PROP1. Furthermore, they showed that maximizing the utilitarian welfare among a set of fair allocations when the number of agents is variable is \myemph{strongly NP-Hard}, while it is NP-Hard when the number of agents is fixed and greater than two. There is a polynomial-time algorithm that determines whether an EF1 allocation exists or not among the utilitarian maximal allocations when there are two agents. Further, they designed \myemph{pseudo-polynomial time} algorithms for both problems with a fixed number of agents.

In light of these open questions in finding efficient EF1 allocations, researchers have explored efficient PROP1 allocations and obtained positive results.

\subsubsection*{\myemph{\textbf{Efficient PROP1 allocations}}}
\paragraph*{\myemph{\textbf{PROP1 + PO}}}

EF1 and PO are satisfied by MNW allocation, which implies that when agents have additive valuations, an allocation that meets PROP1 and PO exists.
~\cite{conitzer2017fairpublicdecisionmaking}.
In public decision settings (a setting that is more general than indivisible goods allocation),~\citeauthor{conitzer2017fairpublicdecisionmaking}\cite{conitzer2017fairpublicdecisionmaking} proved that there is always a PROP1 and PO allocation. They left the complexity of computing such an allocation as an open question.
For goods with additive valuations,~\citeauthor{barman2019PROP1POGoods}\cite{barman2019PROP1POGoods} proposed a \myemph{strongly polynomial-time} algorithm for calculating PROP1 and fPO (stronger than PO) by leveraging \myemph{pure market equilibrium}. In contrast to PO, fPO requires that an allocation is not dominated by fractional (or integral) allocations; fPO is, therefore, a stronger solution concept.
Because finding Pareto improvements while maintaining PROP1 is challenging, they deal with fractional allocations that are improving and convert them into integral allocations. They do so by leveraging fisher markets, i.e., such markets consist of a set of divisible goods, a set of agents with budgets and valuations over these goods.
Fisher markets clear when goods are assigned prices; each agent spends its entire budget on only those most valuable goods per unit spent, and the market is in equilibrium.
From the resource allocation perspective, market equilibrium is applicable since the first welfare theorem states that these equilibria are always Pareto efficient~\cite{mas1995microeconomic}.
In Fisher markets with additive valuations, Eisenberg and Gale's convex program provides an efficient way to determine equilibrium allocations and prices: the primal and dual solutions of the convex program correspond to the equilibrium allocations and prices~\cite{eisenberg1959consensus,roughgarden2010algorithmic}. Strongly polynomial-time algorithms exist for the same~\cite{orlin2010improved,vegh2016strongly}.
The equilibrium allocation is fractionally Pareto efficient (fPO). However, equilibrium allocations are not necessarily integral, i.e., the allocation we get may not be applicable in the context of indivisible goods. Note that (in contrast to computing an arbitrary equilibrium) finding an integral equilibrium is computationally hard, i.e., determining whether a given Fisher market is pure is an NP-hard problem~\cite{barman2019PROP1POGoods}.
In general, agents are assigned fractional shares of goods in a market equilibrium. In other words, we cannot directly incorporate indivisible goods into the market framework.

The authors developed an efficient algorithm that first computes an equilibrium ($A$, $p$) of the market $B$ and, then, round (fractional) allocation $A$ to an integral allocation $A'$ such that it is an integral equilibrium of the market $B'$. The algorithm doesn't change the price of the goods. It defines the new budgets (a bounded change in agents' budgets) to explicitly satisfy budget exhaustion with respect to the computed allocations and the unchanged prices at the end. Since $A'$ is an equilibrium of the Fisher market $B'$, via the first welfare theorem, $A'$ is fPO and PROP1. This conversion to $A'$ is in strongly polynomial time.
Briefly speaking, given a market and its equilibrium $(A,p)$, the spending graph $G(A,p)$ is a forest. The algorithm starts by identifying the root of each tree at some agent in this forest. Then assign child goods to root agent $i$ (with no parents) until adding more child goods to $i$ violates the budget constraint. The remaining child goods are then appropriately assigned to grandchildren agents. After each distribution, delete this parent agent $i$ and all its allocated child goods. Iteratively assign remaining goods to agents until all the goods are allocated. Note that if a good is completely assigned to an agent, i.e., it is integrally assigned, then it will continue to be assigned to the same agent after conversion. 
This algorithm gives weighted-PROP1, which was later mentioned in~\cite{branzei2019choresPROP1PO}. This algorithm satisfies EF$1^1$, weighted-PROP1, and fPO.

\paragraph*{\myemph{PROP1 + USW}}
Among a set of utilitarian maximal allocations,~\citeauthor{azizEF1PROP1plusUSW}~\cite{azizEF1PROP1plusUSW} proved that it is \myemph{strongly NP-hard} to find an allocation that is PROP1. Furthermore, they showed that maximizing the utilitarian welfare among a set of fair allocations when the number of agents is variable is strongly NP-Hard, while it is NP-Hard when the number of agents is fixed and greater than two. There is a polynomial-time algorithm that determines whether a PROP1 allocation exists or not among the utilitarian maximal allocations when there are two agents. Further, they designed pseudo-polynomial time algorithms for both problems with a fixed number of agents.

\paragraph*{\myemph{\textbf{PROPm + PO/USW}}}
Towards finding efficient PROPm allocations, this
area of study remains unexplored, leaving the scope
for future research.

\subsubsection*{\myemph{Efficient EFX allocations}}
\paragraph*{\myemph{EFX + PO}}
In the presence of zero marginal valuations,~\citeauthor{plaut2020EFX}~\cite{plaut2020EFX}
showed that there are additive valuations in
which no EFX allocation is also PO.
In the presence of zero marginal valuations,~\citeauthor{plaut2020EFX}~\cite{plaut2020EFX}
showed that there are general and identical valuations in
which no EFX allocation is also PO. 
In the presence of nonzero marginal valuations, for any number of agents with general but identical valuations, and for two agents with
(possibly distinct) additive valuations, the
leximin++ solution is EFX and PO.
They also argued that the assumption of nonzero marginal utility is quite reasonable with two agents having additive valuations as one agent is indifferent to some good; we can assign the good to the other agent and exclude it from the division process completely. 

\begin{figure}[h]
     \centering
    \includegraphics[width=0.9\textwidth]{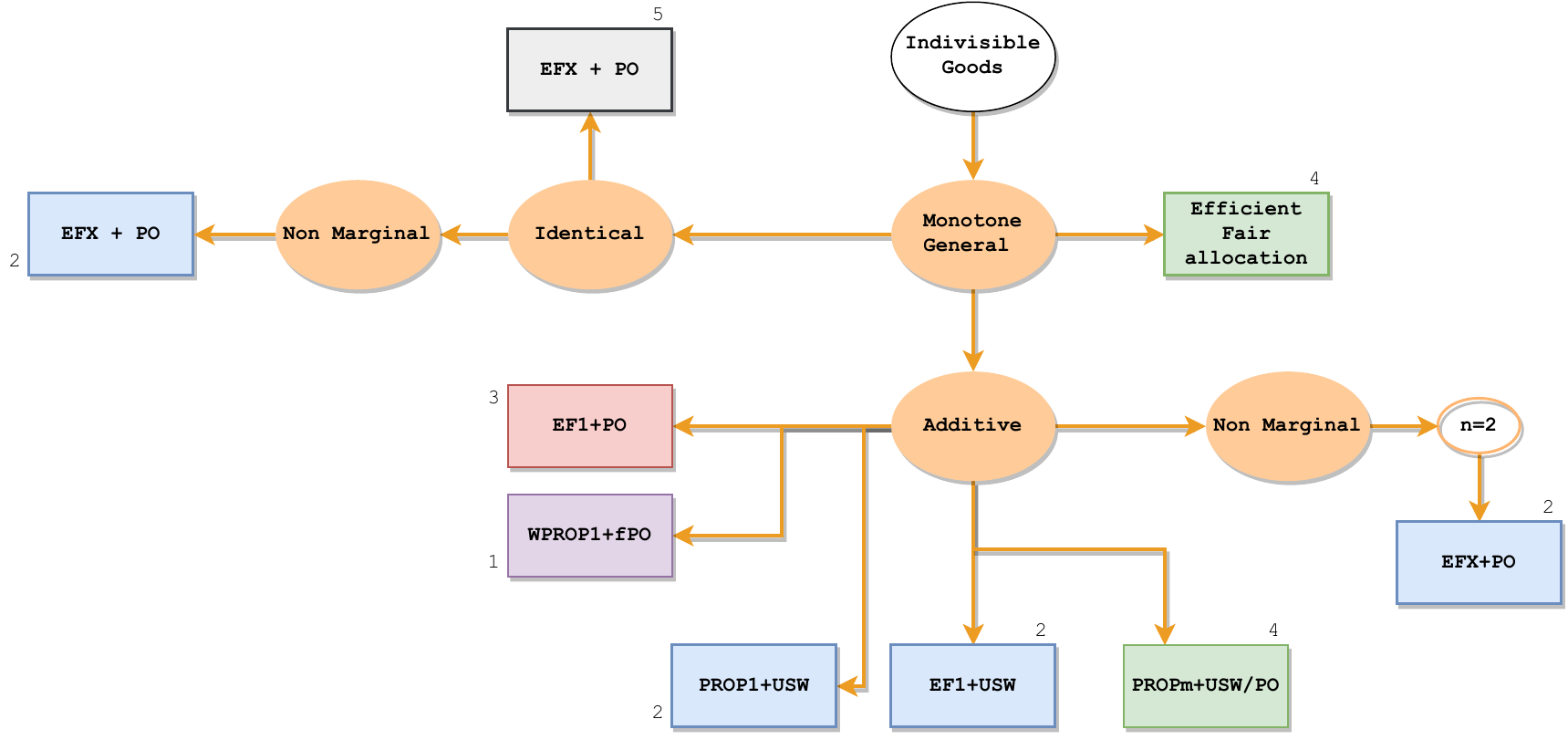}
     \caption{Summary of efficient fairness allocation for indivisible goods. The Violet box (number 1) represents the polynomial-time algorithm. The blue box (number 2) represents exponential time. The grey box (number 5) represents such an allocation does not exist. The red box (number 3) represents pseudo-polynomial time. The green box (number 4) represents open questions.}
     \label{fig:fairandefficientindivisiblegoodssummary}
 \end{figure}

Towards finding efficient, fair allocations, especially for general valuations, this
area of study remains unexplored, leaving the scope
for future research. Moreover, it would be interesting to explore efficient allocation when a fair allocation, i.e., if MMS allocation exists, what are the efficiency guarantees we can achieve?  We summarize our results about the efficient, fair allocation of goods in Figure~\ref{fig:fairandefficientindivisiblegoodssummary}. Next, we move on to the discussion of chores.

\section{Indivisible Chores}
\label{sec:chores}

\subsection{Fair Allocation}
\subsubsection*{\myemph{\textbf{Envy-Freeness:}}}
Similar to indivisible goods, determining the existence of an
envy-free allocation is NP-complete, even in the simple case when agents have binary additive valuations for chores~\cite{bhaskar2020approximateefforindivisiblechores}.


\subsubsection*{\myemph{\textbf{Envy-Freeness up to one item (EF1):}}}
Similarly to goods, a simple round robin gives an EF1 for chores in \myemph{polynomial time} -- O($m n \log m$). The chore division problem might appear to be the ‘opposite’ of the goods problem, and therefore intuitively, one might expect the natural adaptation of algorithms designed to compute an EF1 allocation for goods, to also work for chores. This, however, is not the case. The envy-cycle elimination algorithm \cite{lipton2004approximately} doesn't give EF1 allocation in the case of chores. \citeauthor{bhaskar2020approximateefforindivisiblechores}~\cite{bhaskar2020approximateefforindivisiblechores} presented \myemph{Top-trading envy-cycle elimination algorithm}, a modified Envy-cycle algorithm for chores that gives EF1 for indivisible chores with anti-monotonic valuations. For indivisible goods, resolving arbitrary envy cycles (until the envy graph becomes acyclic) is known to preserve EF1. However, for indivisible chores, the choice of which envy cycle is resolved matters. EF1 for chores entails that any pairwise envy can be addressed by removing some chores from the envious agent’s bundle. The algorithm works by assigning, at each step, an unassigned item with the largest cost to an agent who does not envy anyone else (i.e., a non-envious agent who is a “sink” node in the top-trading envy graph). If the top-trading envy graph does not have a sink, then it must have a cycle. Then resolving the top-trading envy cycles guarantees the existence of a sink agent in the top-trading envy graph.

Now we briefly describe the algorithm for EF1 for chores as follows. In the above algorithm, given a partial allocation $A$, we consider a subgraph of the envy-graph $G_A$ that
we call the top-trading envy graph $T_A$ whose vertices denote the agents, and an edge $(i, k)$ denotes that
agent $i$’s (weakly) most preferred bundle is $A_k$. If the envy graph does not have a sink, then the top-trading envy graph $T_A$ has a cycle. Therefore, resolving top-trading envy cycles guarantees the existence of a sink agent in the envy graph, and it also preserves EF1. This is because every agent involved in the top-trading exchange receives their most preferred bundle after the swap, which means they do not envy anyone else in the next round. 

\vspace{-0.5cm}

\subsubsection*{\myemph{\textbf{Envy-Freeness up to any item (EFX):}}}
An EFX allocation also exists for IDO instances for chores~\cite{li2021wpropxchores}.
As described in~\cite{bhaskar2020approximateefforindivisiblechores}, the Top Trading Envy Cycle algorithm allocates items in any order, and it meets the EF1 requirement. Modifying this algorithm to allocate the largest chores (minimum valued items) to sink agents guarantees EFX allocation for IDO instances. We can ensure that EFX allocation is made for IDO instances if we select the item with the highest cost at each step. This returns an EFX allocation in polynomial time for additive valuations for any IDO instance. For two agents or any agents with IDO valuation functions, EFX allocations always exist.


\paragraph*{\myemph{\textbf{Approximate EFX allocations}}}
While good results have been obtained for the (approximation of) EFX allocations for goods, little is known about how EFX is allocated (approximately). It is not known even if constant approximations of EFX allocations exist for three agents with general additive valuations, let alone whether EFX allocations exist at all.
For indivisible chores, the authors propose algorithms to compute approximate EFX allocations in polynomial time in~\cite{zhou2021approximately}. When there are three agents, the algorithm produces 5-EFX, and when there are four or more agents, it produces $3n^2$, i.e.,$\mathcal{O}(n^2)$ as the approximation ratio. The authors observed that in doing so, i.e., the algorithm in order to allocate approximate EFX often would increase the cost of other agents instead of decreasing the cost of an envied agent, which can lead to inefficient allocation. Another interesting open question in approximate-EFX is, can we get a constant approximate ratio for any number of agents? Apart from approximate-EFX, in the case of goods, EFX with charity is well explored as described in Section; however, it is unknown whether similar results hold for the allocation of chores.


\subsubsection*{\myemph{\textbf{Proportionality (PROP):}}}
Similar to indivisible goods, there may not be a proportional allocation for indivisible chores. Determining if a proportional allocation is possible, even for just two agents, is a problem that is NP-Complete~\cite{bouveret2016characterizingconflictsinFD}.

\begin{tcolorbox}[enhanced,drop fuzzy shadow southwest, colframe=red!20!white!30!black,colback=LimeGreen!10]
\begin{algorithm}[H]
\caption{\myemph{\myemph{EF1 for Chores with General Valuations}}}\label{algo:ef1chores}
\begin{algorithmic}[1]
\State Initialize $A \leftarrow (\emptyset,\emptyset,\ldots,\emptyset)$
\For{$k \leftarrow 1 \textit{ to } m$}
\If{there is no sink agent in envy graph $G_A$}
\State $C \leftarrow $ any cycle in $T_A$
\State $A \leftarrow A^C$
\EndIf
\State Chose a sink $j$ in the envy graph $G_A$
\State $A_j \leftarrow A_j \cup k$
\EndFor
\Return Allocation $A$
\end{algorithmic}
\end{algorithm}
\end{tcolorbox}

 \begin{figure}[h]
     \centering
    \includegraphics[width=0.9\textwidth]{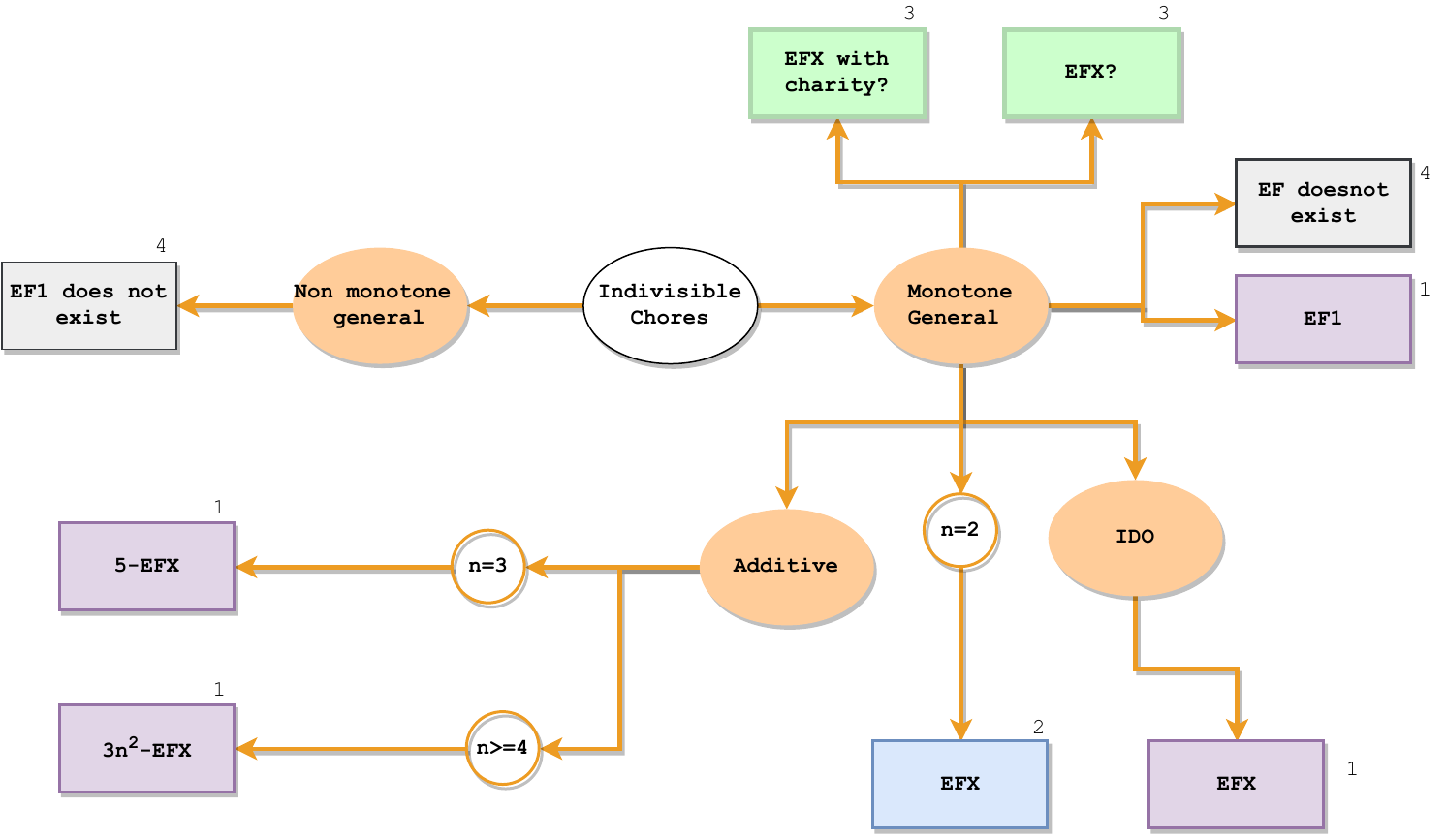}
     \caption{Summary of EF fairness notion for indivisible chores. The Violet box (number 1) represents the polynomial-time algorithm. The blue box (number 2) represents exponential time. The grey box (number 4) represents such an allocation does not exist. The green box (number 3) represents an open question.}
     \label{fig:EFindivisiblechoressummary}
 \end{figure}

\subsubsection*{\myemph{\textbf{Proportionality up to one item (PROP1):}}}
When valuations are additive, EF1 and PROP1 are equivalent. This means that any algorithm that results in EF1 allocations also satisfies PROP1. According to a study by Aziz et al. (\cite{aziz2022fairindivisiblegoodsandchores}), using the PROP1 concept allows for not only fairness but also connectivity in the allocation of mixed goods (including chores) in a contiguous and proportional manner and can be computed efficiently.


\subsubsection*{\myemph{\textbf{Proportionality up to any item (PROPX):}}}
It is widely acknowledged that EFX (Envyfreeness up to any item) is stronger to PROPX (Proportionality up to any item) when it comes to the allocation of chores. This stands in contrast to the allocation of goods, where PROPX has been found to be more stronger. In their study,~\cite{li2021wpropxchores} examined the concept of weighted proportionality up to any item and established that a weighted PROPX allocation exists and can be calculated in a computationally efficient manner. Therefore, while PROPX may not be applicable to indivisible goods, it can be calculated efficiently for indivisible chores. 
The \myemph{``Bid-and-Take''} algorithm, a variation of the "Top trading envy cycle algorithm" \cite{bhaskar2020approximateefforindivisiblechores}, is presented by~\cite{li2021wpropxchores}. In this algorithm, they choose the largest tasks (items with the lowest value) and assign them to the active agent who has the lowest current social cost for the item. Initially, all agents are active. When an agent's cumulative cost exceeds their proportional share, they are deactivated. It is important to note that in \cite{bhaskar2020approximateefforindivisiblechores}, items were allocated in any order and were EF1. However, this modification guarantees that the allocation is EFX (and therefore PROPX) for IDO instances. 

The authors proposed using a similar approach to algorithms for approximate MMS, where they reduce the instance to IDO to find the $\alpha$-MMS and then reconvert it back to the original instance, while still maintaining its $\alpha$-MMS status. They state that if a polynomial time algorithm exists that can compute an $\alpha$-WPROPX allocation for any IDO instance, then there also exists a polynomial time algorithm that can compute an $\alpha$-WPROPX allocation for any additive instance.
This modified envy-cycle elimination algorithm ensures both weighted PROPX and 4/3 approximate MMS simultaneously. Additionally, the authors have also found that an EFX and a weighted EF1 allocation for indivisible chores exist if all agents have the same ordinal preference.
It should be noted that while all PROPX allocations are PROP1, the reverse is not necessarily true. As demonstrated for indivisible chores, wPROPX allocations can be computed in a \myemph{polynomial time}, the emphasis is placed on wPROPX allocations, and any findings apply to PROP1.

 \begin{figure}[h]
     \centering
    \includegraphics[width=0.5\textwidth]{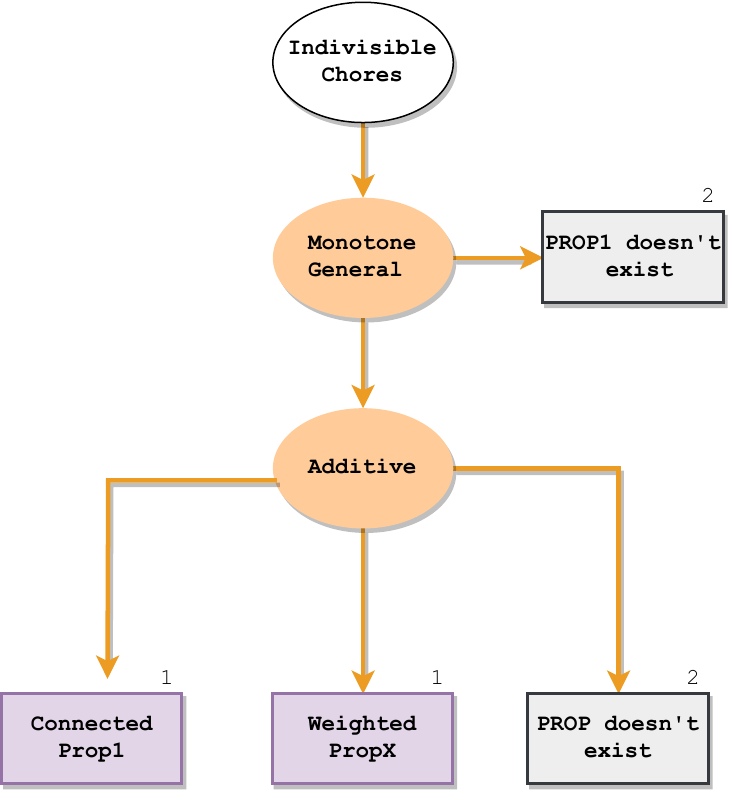}
     \caption{Summary of PROP fairness notion for indivisible chores. The Violet box (number 1) represents the polynomial-time algorithm. The grey box (number 2) represents such an allocation does not exist.}
     \label{fig:PROPindivisiblechoressummary}
 \end{figure}

\subsubsection*{\myemph{\textbf{Maximin Share Allocation (MMS):}}}
The problem of allocating MMS allocation to chores is closely related to \myemph{bin packing problem/job scheduling problem}. 

\citeauthor{azizmmschores}~\cite{azizmmschores} presented an example based on~\cite{procaccia2014mmsdoesntexist} with three agents with additive valuations where every allocation results in some agents having a lower MMS value than they should. There is no guarantee that an MMS allocation for chores exists, even with additive valuations, as demonstrated in~\cite{azizmmschores,feige2021tightnegativeexampleformms}. 
While it has been shown that MMS allocations are not guaranteed to exist for indivisible
chores~\cite{azizmmschores}, there are many works that study its approximations~\cite{azizmmschores,barmanapproxmms2020,huang2021algorithmic}. The state-of-the-art approximation ratio for MMS allocation for indivisible chores is 11/9~\cite{huang2021algorithmic}

\begin{theorem} \cite{azizmmschores}
Given a chore allocation problem $(N,M,\mathcal{V})$, for any $n \ge 3$, there exists $M$ and (additive) valuations that do not admit an MMS allocation.
\end{theorem}

\vspace{0.5cm}

\subsubsection*{\myemph{\textbf{Approximately MMS Allocation ($\alpha$-MMS):}}}

Researchers have been working on finding efficient algorithms for approximating the maximum minimum share (MMS) allocation, particularly for situations where MMS does not always exist, such as in the case of chores. One such algorithm for 2-MMS has been demonstrated to be polynomial time by \citeauthor{azizmmschores}\cite{azizmmschores}. This result has been improved upon by \citeauthor{barmanapproxmms2020}, who presented a polynomial time algorithm for 4/3-MMS. In the latest development, \citeauthor{huang2021algorithmic} showed that an \myemph{11/9-MMS} allocation could be achieved by converting the instances to identical ordinal instances (IDO) and then solving for approximate maximin allocation. The authors proposed a mechanism similar to the \myemph{First Fit Decreasing (FFD) algorithm} for bin packing problems to allocate chores. This involves starting with high-valued chores and adding them to the bundle in decreasing order until an agent's threshold is met, and then allocating that bundle to that agent. The algorithm repeats this process for all agents until all chores are allocated. The authors also provided a polynomial time approximation scheme (PTAS) and a polynomial time algorithm for 5/4-MMS allocation.

To compute the 11/9-MMS allocation, the algorithm first converts the instance to an IDO instance and orders the chores in decreasing order. Then, starting with high-valued chores, the algorithm adds chores to the bundle until an agent's threshold is met and allocates the bundle to that agent. This process is repeated for all agents until all chores are allocated. The authors proved that if the threshold value is set to 11/9, then the algorithm terminates with an 11/9-MMS allocation, with all chores being allocated. However, the best approximation achievable using this method is an open question.

In summary, researchers have been exploring ways to approximate the MMS allocation for situations where it does not always exist, such as in the case of chores. The latest development is an algorithm that can achieve an \myemph{11/9-MMS} allocation by converting instances to IDO instances and using a mechanism similar to the FFD algorithm for allocating chores. This algorithm has been shown to be polynomial time, and the authors also provided a PTAS and a polynomial time algorithm for \myemph{5/4-MMS} allocation. However, the best approximation achievable using this method is still unknown.

\vspace{-1.5cm}

\begin{tcolorbox}[enhanced,drop fuzzy shadow southwest, colframe=red!20!white!30!black,colback=LimeGreen!10]
\begin{algorithm}[H]
\caption{\myemph{Algorithm for $11/9$-MMS}}\label{algo:alphammschores}
\begin{algorithmic}[1]
\State Assume IDO instance and threshold values of each agent $(\alpha_1,\alpha_2,\ldots,\alpha_3)$
\State Assume $\forall i, \mu_i = 1$
\For{$i \leftarrow 1 \textit{ to } n$}
\State $K \leftarrow \emptyset$
\For{$k \leftarrow 1 \textit{ to } m$}
\If{$j \exists N, \; v_j(A_j \cup M_k) \le \alpha_j $}
\State $K \leftarrow K \cup M_k$
\EndIf
\EndFor
\State $M \leftarrow M \setminus K$
\State Let $j$ be the agent who is allocated K, i.e., $v_j(K)\le \alpha_j$
\State $A_j \leftarrow K$ and $N \leftarrow N \setminus j$
\EndFor
\Return Allocation $A$
\end{algorithmic}
\end{algorithm}
\end{tcolorbox}

 \begin{figure}[H]
     \centering
    \includegraphics[width=0.8\textwidth]{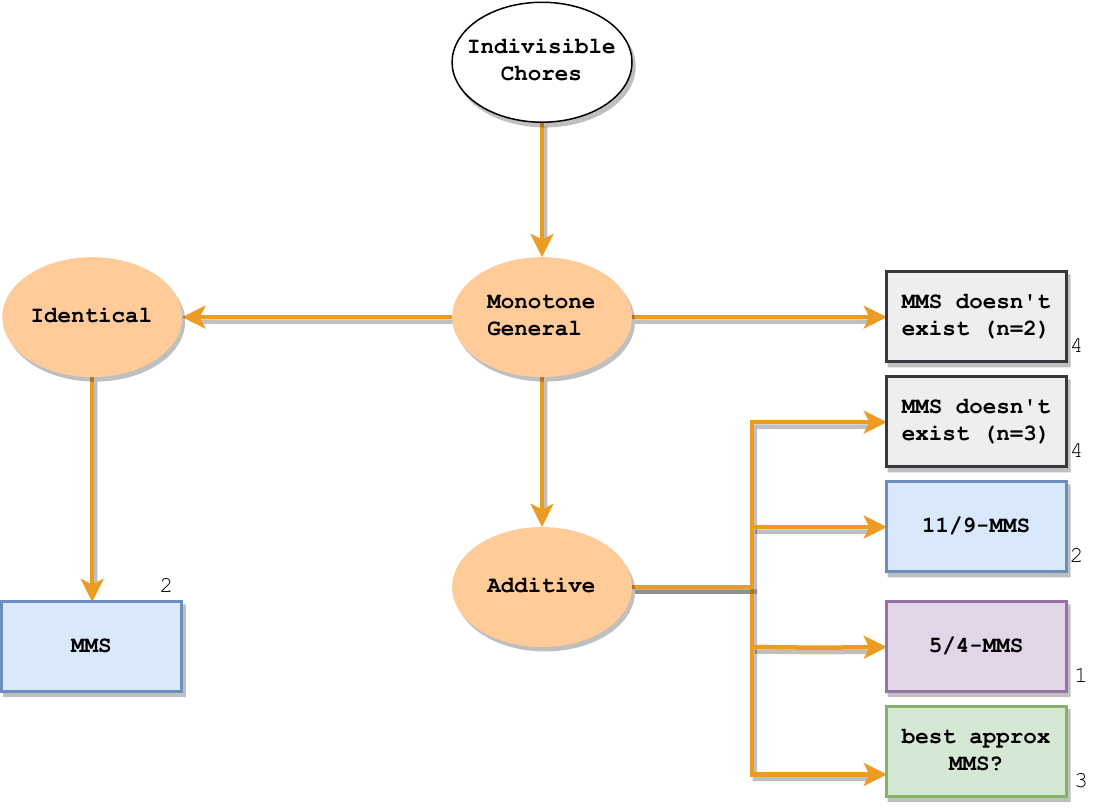}
     \caption{Summary of MMS fairness notion for indivisible chores. The Violet box (number 1) represents the polynomial-time algorithm. The blue box (number 2) represents the exponential-time algorithm. The green box (number 3) represents an open question. The grey box (number 4) represents such an allocation does not exist.}
     \label{fig:MMSindivisiblechoressummary}
 \end{figure}

\subsection{Fair and Efficient Allocation}
\subsubsection*{\myemph{\textbf{Efficient EF1 allocations}}}

When valuations are negative, the existence and complexity of an allocation that satisfies EF1 (envy-freeness up to one item) and PO (proportional) properties is still an \myemph{open problem}. However, alternative fairness criteria, such as \myemph{EF$^1_1$ (envy-freeness up to one item and one penny)} introduced in \cite{branzei2019choresPROP1PO}, do exist and can be efficiently computed along with fPO (fractional proportional) allocation. Additionally, the problem of maximizing utilitarian welfare while ensuring EF1 allocation is not fully explored in chore allocation.

In summary, despite the challenge posed by negative valuations, there exist fair allocation criteria that can be efficiently computed, and there is still room for further research in maximizing welfare while satisfying fairness constraints in the context of chore allocation.

\subsubsection*{\myemph{\textbf{Efficient PROP1 allocations}}}

In a recent study, Branzei et al. \cite{branzei2019choresPROP1PO} demonstrated the existence of a \myemph{strongly polynomial-time} algorithm for solving the chores problem. This algorithm guarantees a weighted PROP1 and PO allocation, provided the number of agents or items is fixed. The underlying algorithm used in their study is similar to the pure market equilibria approach proposed by Barman et al. \cite{barman2019PROP1POGoods}.

The researchers also established that for any chore division problem, there exists an indivisible allocation $A$ that is Pareto optimal in the divisible problem and satisfies weighted-EF$1^1$ and weighted-Prop1. Furthermore, they computed a competitive allocation $A'$ and demonstrated that an integral competitive allocation $A$ with a budget satisfying could be constructed from $A'$. This allocation is Pareto optimal by the first welfare theorem.

In summary, Branzei et al. proposed an efficient algorithm for solving the chores problem and provided theoretical guarantees for the Pareto optimality and efficiency of their proposed allocation. Their findings provide valuable insights into the design of allocation algorithms for resource allocation problems. It is important to note that the existence of PROPX and PO allocation for chore allocation problems remains an \myemph{open question}.

 \begin{figure}[h]
     \centering
    \includegraphics[width=0.8\textwidth]{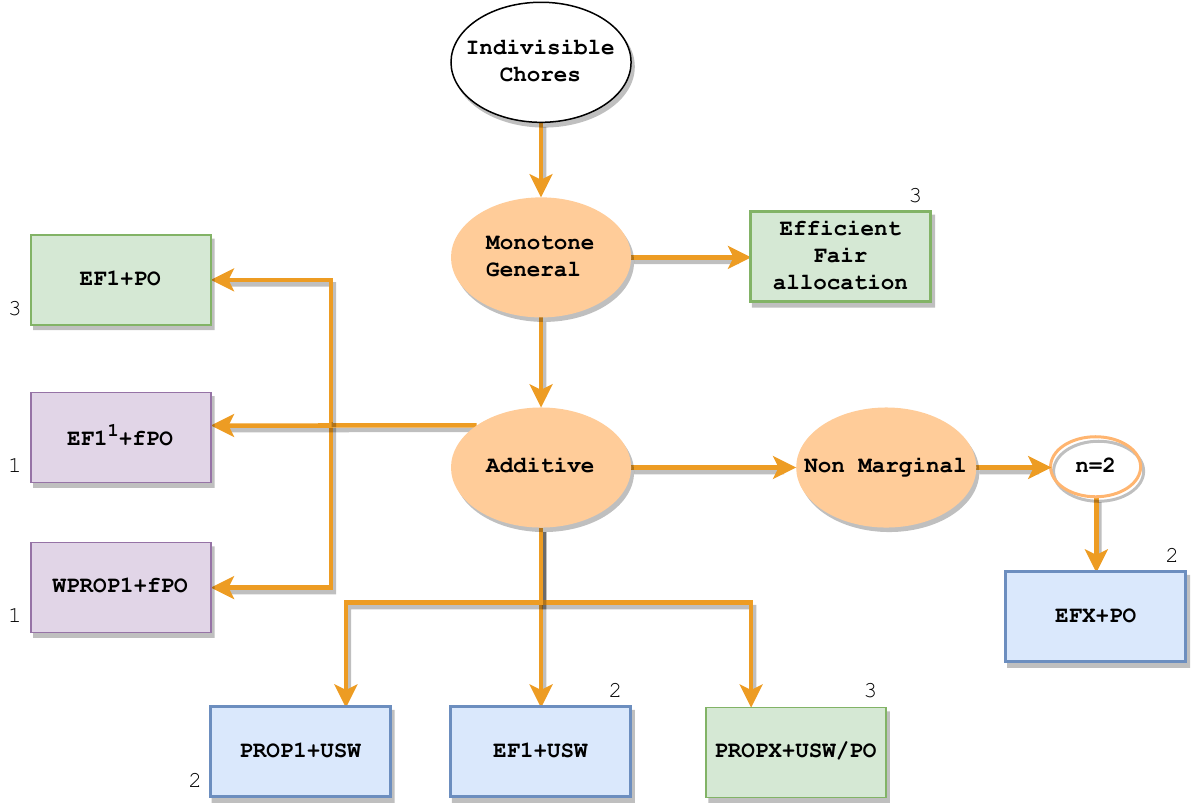}
     \caption{Summary of efficient fairness notion for indivisible chores. The Violet box (number 1) represents the polynomial-time algorithm. The blue box (number 2) represents exponential time. The green box (number 3) represents an open question.}
     \label{fig:efficientindivisiblechoressummary}
 \end{figure}

We summarize the results discussed in this section in the following Table~\ref{tab:summarytable1}. The red line represents impossibilities. And brown line represents open questions in the literature. Even though there is a rich literature, there are still a lot of open problems that need to be resolved.


\begin{table}[!t]
\centering
\caption{\label{tab:summarytable1}Summary of Fair Division for Indivisible Goods or/and Chores}
\resizebox{16cm}{!}{
\begin{tabular}{|c|c|c|c|c|c|c|c|c|}
\hline
\rowcolor[HTML]{b7df86}
\textbf{{Fairness}} & \textbf{{Efficiency}} & \textbf{{Agents}} & \textbf{{Items}} & \textbf{{Valuations}} & 
\textbf{{Existence}} & 
\textbf{{Computation}} & \textbf{{Algorithm}} & \textbf{{Paper}} \\ \hline
\rowcolor[HTML]{FFE4E1}
EF & any & any & any & any & no & NP-Hard & - & - 
\\ \hline
\multirow{3}{*}{EF1} & \multirow{3}{*}{-} & \multirow{3}{*}{any} & goods & \multirow{3}{*}{additive}  & \multirow{3}{*}{yes} & \multirow{3}{*}{polynomial} & \multirow{2}{*}{round robin} & \tiny{\citeauthor{caragiannis2019unreasonable}~\cite{caragiannis2019unreasonable}} \\ \cline{4-4}\cline{9-9}
& & & chores &  &  & &  & \multirow{2}{*}{ 
\tiny{
\citeauthor{aziz2022fairindivisiblegoodsandchores}~\cite{aziz2022fairindivisiblegoodsandchores}}} \\ \cline{4-4}\cline{8-8}
& & & comb &  &  & & 
\begin{tabular}[c]{@{}l@{}}doubly round\\robin elimination\end{tabular}
 &   \\ \hline
 \multirow{3}{*}{EF1} & \multirow{3}{*}{-} & \multirow{3}{*}{any} & goods & \multirow{3}{*}{general} & \multirow{3}{*}{yes} & \multirow{3}{*}{polynomial} & 
\begin{tabular}[c]{@{}l@{}}envy cycle\\elimination\end{tabular}
& \tiny{\citeauthor{lipton2004approximately}~\cite{lipton2004approximately}} \\ \cline{4-4}\cline{8-9}
& & & chores &  &  & &\multirow{2}{*}{
\begin{tabular}[c]{@{}l@{}}modified envy\\cycle elimination\end{tabular}}  & \multirow{2}{*}{\tiny{\citeauthor{bhaskar2020approximateefforindivisiblechores}~\cite{bhaskar2020approximateefforindivisiblechores}}}\\ \cline{4-4}
& & & comb &  &  &  &  &\\ \hline
\rowcolor[HTML]{dbefc3}
EF1 & - & any & any & arbitrary & ? & ? & ? & ?  \\ \hline
 \rowcolor[HTML]{dbefc3}
EF1 & PO & any & any & general & ? & ? & ? & ? \\ \hline
 &  &  & goods &  & yes & \begin{tabular}[c]{@{}l@{}}pseudo\\ polynomial\end{tabular}& market equilibria & 
\begin{tabular}[c]{@{}l@{}}\tiny{\citeauthor{barman2018findingEF1PO}
}\\\tiny{\cite{barman2018findingEF1PO}}\end{tabular}
 
 \\ \cline{4-4} \cline{6-9} 
 \rowcolor[HTML]{dbefc3}
 &  &  & chores &  &  &  &  &  \\ \cline{4-4}
  \rowcolor[HTML]{dbefc3}
\multirow{-3}{*}{EF1} & \multirow{-3}{*}{PO} & \multirow{-3}{*}{any} & comb & \multirow{-3}{*}{additive} & \multirow{-2}{*}{?} & \multirow{-2}{*}{?} & \multirow{-2}{*}{?} & \multirow{-2}{*}{?} \\ \hline
EF1 & USW & any & any & any & yes & NP-Hard & ? & ? \\ \hline
EF1 & USW & any & good & additive & yes & exponential & dynamic programming & \tiny{\citeauthor{azizEF1PROP1plusUSW}~\cite{azizEF1PROP1plusUSW}} \\ \hline
 \rowcolor[HTML]{dbefc3}
 EFX & any & any & \begin{tabular}[c]{@{}l@{}}goods/\\chores\end{tabular} & general & ? & ? & ? & ? \\ \hline
 \rowcolor[HTML]{dbefc3}
 EFX & any & any & \begin{tabular}[c]{@{}l@{}}goods/\\chores\end{tabular} & additive & ? & ? & ? & ?  \\ \hline
 \rowcolor[HTML]{FFE4E1}
 EFX & - & any & comb & 
  \begin{tabular}[c]{@{}l@{}}lexicographic sub\\ domain of additive\end{tabular}
   & no &  &  & \tiny{\citeauthor{hosseini2022fairly}~\cite{hosseini2022fairly}} \\ \hline
  &  &  & goods &  & yes & exponential & leximin++ & \tiny{\citeauthor{plaut2020EFX}~\cite{plaut2020EFX}} \\
  \cline{4-4} \cline{6-9} 
  \rowcolor[HTML]{dbefc3}
  &  &  & chores &  &  &  &  &  \\ \cline{4-4}
  \rowcolor[HTML]{dbefc3}
 \multirow{-3}{*}{EFX} & \multirow{-3}{*}{-} & \multirow{-3}{*}{any} & comb & \multirow{-3}{*}{IDO general} & \multirow{-2}{*}{?} & \multirow{-2}{*}{?} & \multirow{-2}{*}{?} & \multirow{-2}{*}{?} \\ \hline
  &  &  & goods &  &  &  & envy cycle & \tiny{\citeauthor{plaut2020EFX}~\cite{plaut2020EFX}} \\ \cline{4-4} \cline{8-9} 
 &  &  & chores &  & \multirow{-2}{*}{yes} & \multirow{-2}{*}{polynomial} & envy cycle & \tiny{\citeauthor{li2021wpropxchores}~\cite{li2021wpropxchores}} \\ \cline{4-4} \cline{6-9} 
  \rowcolor[HTML]{dbefc3}
\multirow{-3}{*}{EFX} & \multirow{-3}{*}{-} & \multirow{-3}{*}{any} & comb &
\multirow{-3}{*}{IDO additive} & ? & ? & ? & ? \\ \hline
EFX & - & any  & goods & 2-value instances & yes & polynomial & match\&freeze & \tiny{\citeauthor{amanatidis2021mnwandEFXstories}~\cite{amanatidis2021mnwandEFXstories}} \\ \hline
EFX &  &  & goods & \begin{tabular}[c]{@{}l@{}}interval valuation\\ $[v_i - 2v_i]$\end{tabular} & yes & polynomial & modified RR & \tiny{\citeauthor{amanatidis2021mnwandEFXstories}~\cite{amanatidis2021mnwandEFXstories}} \\ \hline
EFX & - & $n=3$ & goods & additive & yes & \begin{tabular}[c]{@{}l@{}}pseudo\\ polynomial\end{tabular} &  & \tiny{\citeauthor{chaudhury2020efx}~\cite{chaudhury2020efx}} \\ \hline
 & - &  & goods &  &  &  & 
 \begin{tabular}[c]{@{}l@{}} cut-choose,\\leximin++\end{tabular}
  & \tiny{\citeauthor{plaut2020EFX}~\cite{plaut2020EFX}} \\ \cline{4-4} \cline{8-9} 
 & - &   & chores &  &  &  & cut-choose &  ~\tiny{ Implied \citeauthor{zhou2021approximately}~\cite{zhou2021approximately}} \\ \cline{4-4} \cline{8-9} 
\multirow{-3}{*}{EFX} & \multirow{-3}{*}{-} & \multirow{-3}{*}{$n=2$} & comb & \multirow{-3}{*}{general} & \multirow{-3}{*}{yes} & \multirow{-3}{*}{exponential} & cut-choose & \tiny{Implied} \\ \hline
EFX & - & $n=2$ & goods & additive &  & polynomial & envy cycle & \tiny{\citeauthor{plaut2020EFX}~\cite{plaut2020EFX}} \\ \hline
1/2 EFX & - &  any & goods & subadditive & yes & polynomial & envy cycle & \tiny{\citeauthor{plaut2020EFX}~\cite{plaut2020EFX}} \\ \hline
0.618 EFX & - & any & goods & additive & yes &  polynomial &  & \tiny{\citeauthor{amanatidis2020multiplebirds}~\cite{amanatidis2020multiplebirds}} \\ \hline
\begin{tabular}[c]{@{}l@{}}EFX with\\charity\end{tabular}
& - & any & goods & additive & yes  & polynomial &  & \tiny{\citeauthor{caragiannis2019envyDonation}~\cite{caragiannis2019envyDonation}} \\ \hline
\begin{tabular}[c]{@{}l@{}}approx EFX\\ with charity\end{tabular}
  & - & any & goods & additive & yes  & polynomial &  & \tiny{\citeauthor{chaudhury2021improving}~\cite{chaudhury2021improving}} \\ \hline
  \rowcolor[HTML]{dbefc3}
approx EFX & any & any & \begin{tabular}[c]{@{}l@{}}chores/\\comb\end{tabular} & any & ? & ? & ? & ? \\ \hline
  \rowcolor[HTML]{dbefc3}
\begin{tabular}[c]{@{}l@{}}EFX with\\charity\end{tabular}& any & any & \begin{tabular}[c]{@{}l@{}}chores/\\comb\end{tabular} & any & ? & ? & ? & ? \\ \hline
\multicolumn{4}{r}{\textit{Continued on next page}}\\
\end{tabular}}
\end{table}

\newpage
\begin{table}[!t]
\centering
\caption*{}
\resizebox{16cm}{!}{
\begin{tabular}{|c|c|c|c|c|c|c|c|c|}
\hline
\rowcolor[HTML]{b7df86}
\textbf{{Fairness}} & \textbf{{Efficiency}} & \textbf{{Agents}} & \textbf{{Items}} & \textbf{{Valuations}} & 
\textbf{{Existence}} & 
\textbf{{Computation}} & \textbf{{Algorithm}} & \textbf{{Paper}} \\ \hline
EFX & PO & $n=2$ & goods & \begin{tabular}[c]{@{}l@{}}general nonzero\\ marginal utility\end{tabular} &  & exponential & \begin{tabular}[c]{@{}l@{}} cut-choose,\\leximin++\end{tabular} & \tiny{\citeauthor{plaut2020EFX}~\cite{plaut2020EFX}} \\ \hline
EFX & PO & any  & goods & \begin{tabular}[c]{@{}l@{}}identical general\\ non zero marginal\end{tabular} &  & exponential & leximin++ & \tiny{\citeauthor{plaut2020EFX}~\cite{plaut2020EFX}} \\ \hline
EFX & PO & any & goods & 2-value instances &  & exponential & MNW & \tiny{\citeauthor{amanatidis2021mnwandEFXstories}~\cite{amanatidis2021mnwandEFXstories}} \\ \hline
\begin{tabular}[c]{@{}l@{}}EFX, PROPX,\\4/3 MMS\end{tabular}&  &  & chores & IDO additive &  & polynomial & envy cycle & \tiny{\citeauthor{li2021wpropxchores}~\cite{li2021wpropxchores}} 
 \\ \hline
 \rowcolor[HTML]{dbefc3}
 PROP & \multicolumn{1}{c|}{any} & any & any & any & no & NP-Hard & ? & ? \\ \hline
 \multirow{3}{*}{PROP1} & \multicolumn{1}{c|}{\multirow{3}{*}{-}} & \multirow{3}{*}{any} & goods & \multirow{3}{*}{additive} & \multirow{3}{*}{yes} & \multirow{3}{*}{polynomial} & \multirow{3}{*}{\begin{tabular}[c]{@{}l@{}}EF1 implies PROP1\\ for additive\end{tabular}} & \multirow{3}{*}
 \tiny{
 { \begin{tabular}[c]{@{}l@{}}\citeauthor{caragiannis2019unreasonable}~\cite{caragiannis2019unreasonable} \\  \citeauthor{aziz2022fairindivisiblegoodsandchores}~\cite{aziz2022fairindivisiblegoodsandchores} \end{tabular}
}} \\ \cline{4-4}
 & \multicolumn{1}{c|}{} &  & chores &  &  &  &  &  \\ \cline{4-4}
 & \multicolumn{1}{c|}{} &  & comb &  &  &  &  &  \\ \hline 
\begin{tabular}[c]{@{}l@{}}connected\\PROP1\end{tabular}
  & \multicolumn{1}{c|}{-} & any & any & additive & yes & polynomial &  & \tiny{\citeauthor{aziz2022fairindivisiblegoodsandchores}~\cite{aziz2022fairindivisiblegoodsandchores}} \\ \hline
wPROP1 & \multicolumn{1}{c|}{fPO} & any & any & additive & yes & polynomial & market equilibria & \tiny{\citeauthor{aziz2020polynomialPROP1PO}~\cite{aziz2020polynomialPROP1PO}} \\ \hline
\multirow{2}{*}{PROP1} & \multirow{2}{*}{USW} & \multirow{2}{*}{any} & goods & \multirow{2}{*}{additive} & \multirow{2}{*}{yes} & exponential & 
\begin{tabular}[c]{@{}l@{}}dynamic  \\ programming \end{tabular}
 & \tiny{\citeauthor{azizEF1PROP1plusUSW}~\cite{azizEF1PROP1plusUSW}} \\ \cline{4-4} \cline{7-9} 
\rowcolor[HTML]{dbefc3}
 &  &  & any &  &  & ? & ? & ? \\ \hline
 \rowcolor[HTML]{FFE4E1} \begin{tabular}[c]{@{}l@{}}PROPX\\wPROPX\end{tabular}
& \multicolumn{1}{c|}{-} & any & goods & additive & no & - & - & \tiny{\citeauthor{aziz2022fairindivisiblegoodsandchores}~\cite{aziz2022fairindivisiblegoodsandchores}} \\ \hline
 & \multicolumn{1}{c|}{} &  & comb &  &  &  &  & Implied \\ \hline
 & \multicolumn{1}{c|}{} &  & chores &  &  & polynomial & envy cycle, bid and take & \tiny{\citeauthor{li2021wpropxchores}~\cite{li2021wpropxchores}} \\ \hline
  \rowcolor[HTML]{dbefc3}
PROPX & \multicolumn{1}{c|}{PO} & any & chores & additive & ? & ? & ? & ? \\ \hline
 \rowcolor[HTML]{dbefc3}
PROPX & \multicolumn{1}{c|}{USW} & any & chores & additive & yes & NP-Hard & ? & ? \\ \hline
PROPm & \multicolumn{1}{c|}{-} & any & goods & additive & yes & polynomial & \begin{tabular}[c]{@{}l@{}}divide and conquer \\ plus graph reassigning\end{tabular} & \tiny{\citeauthor{baklanov2021propmfinalllyy}~\cite{baklanov2021propmfinalllyy}} \\ \hline
 \rowcolor[HTML]{dbefc3}
PROPm & \multicolumn{1}{c|}{} &  & comb & additive & ? & ? & ? & ? \\ \hline
 \rowcolor[HTML]{dbefc3}
PROPm & \multicolumn{1}{c|}{PO} & any & any & additive & ? & ? & ? & ? \\ \hline
 \rowcolor[HTML]{dbefc3}
PROP & \multicolumn{1}{c|}{USW} & any & any & additive & yes & NP-Hard & ? & ? \\ \hline
\rowcolor[HTML]{FFE4E1}
MMS & - & any & any & additive & no & NP-Hard & - & 
\tiny{
\begin{tabular}[c]{@{}l@{}}\citeauthor{procaccia2014mmsdoesntexist}~\cite{procaccia2014mmsdoesntexist} \\  \tiny{\citeauthor{azizmmschores}~\cite{azizmmschores}} \end{tabular}} \\ \hline
MMS & - & any & any & IDO & yes & NP-Hard &  & 
\tiny{\citeauthor{bouveret2016characterizingconflictsinFD}~\cite{bouveret2016characterizingconflictsinFD}} \\ \hline
MMS & - & any & goods & 0,1 & yes & polynomial & RR with modification & \tiny{\citeauthor{bouveret2016characterizingconflictsinFD}~\cite{bouveret2016characterizingconflictsinFD}} \\ \hline
MMS & - & any & goods & 0,1,2 & yes & polynomial & RR with modification & \tiny{\citeauthor{Amanatidismms2017}~\cite{Amanatidismms2017}} \\ \hline
MMS & - & $m \le n+3$ & goods & additive & yes &  &  & \tiny{\citeauthor{bouveret2016characterizingconflictsinFD}~\cite{bouveret2016characterizingconflictsinFD}} \\ \hline
MMS & - & $n=2$ & goods & additive & yes & exponential & cut and choose & \tiny{\citeauthor{bouveret2016characterizingconflictsinFD}~\cite{bouveret2016characterizingconflictsinFD}} \\ \hline
3/4 MMS & - & any & goods & additive & yes & polynomial & \multirow{2}{*}{
\begin{tabular}[c]{@{}l@{}}valid reduction,\\clustering, bag filling\end{tabular}} & \tiny{\citeauthor{garg2021improved}~\cite{garg2021improved}} \\ \cline{1-7} \cline{9-9} 
3/4+1/12n & - & any & goods & additive & yes & exponential &  & \tiny{\citeauthor{garg2021improved}~\cite{garg2021improved}} \\ \hline
11/9 MMS & - & any & chores & additive & yes & exponential & \multirow{2}{*}{
\begin{tabular}[c]{@{}l@{}}modified bag filling\\ with threshold\end{tabular}} & \tiny{\citeauthor{huang2021algorithmic}~\cite{huang2021algorithmic}} \\ \cline{1-7} \cline{9-9} 
5/4 MMS & - & any & chores & additive & yes & polynomial &  & \tiny{\citeauthor{huang2021algorithmic}~\cite{huang2021algorithmic}} \\ \hline
\rowcolor[HTML]{FFE4E1}
$\alpha$-MMS & any & any & comb & additive & no & - & - & \tiny{\citeauthor{kulkarni2021indivisible}~\cite{kulkarni2021indivisible}} \\ \hline
\rowcolor[HTML]{dbefc3}
$\alpha$-MMS & PO & any & any & additive & ? & ? & ? & ? \\ \hline
\rowcolor[HTML]{dbefc3}
$\alpha$-MMS & USW & any & any & any & yes & ? & ? & ? \\ \hline
\end{tabular}
}
\end{table}

\newpage

\section{Combination of Indivisible Goods and Chores}
\label{sec:combination}

\subsection{Fair Allocation}
\label{subsec:comb_fair}
\subsubsection*{\myemph{\textbf{Envy-freeness up to one item (EF1):}}}
\citeauthor{aziz2022fairindivisiblegoodsandchores} proposed a \myemph{double round-robin} algorithm for the combination of goods and chores when agents have additive valuations for computing EF1 allocation in polynomial time. The algorithm involves two phases and applies the round-robin method twice, once in a clockwise direction and once in an anticlockwise direction.
During the first phase, the algorithm allocates chores to agents based on their non-positive utility using the round-robin algorithm. In the second phase, the algorithm allocates the remaining goods to agents using the reversed round-robin algorithm, starting with the agent who chose the last in the first phase.

For more general valuations, \citeauthor{bhaskar2020approximateefforindivisiblechores} presented a modified version of the \myemph{envy cycle elimination}, which allocates goods and chores to agents in a doubly monotone instance where the items are indivisible. The algorithm runs in two phases. In the first phase, the algorithm allocates items that are goods for at least one agent using the envy-cycle elimination algorithm \cite{lipton2004approximately}. However, this is done only for the subgraph of agents who consider the item a good. In the second phase, the algorithm allocates items that are chores to the remaining agents using the top-trading envy cycle elimination algorithm. The paper's authors have proven that the modified algorithm returns an EF1 (envy-free up to one item) allocation. Additionally, they maintain the invariant that the partial allocation remains EF1 at every algorithm step. Moreover, the algorithm terminates in \myemph{polynomial time}.


\subsubsection*{\myemph{\textbf{Envy-freeness up to any item (EFX):}}}

\citeauthor{hosseini2020fair} investigated the problem of finding EFX allocation in the presence of \myemph{lexicographic valuations}, a subclass of additive valuations, where items can be goods or chores. The authors demonstrated that an envy-free allocation existence is computationally hard for instances where only chores are considered under lexicographic preferences, in contrast to the polynomial-time solvability of the goods-only case. They also showed that even in the case of combinations of goods and chores, an EFX allocation may not exist. Despite these challenges, the authors identified a domain restriction where EFX allocations are guaranteed to exist and can be computed efficiently. Furthermore, they showed that a different notion of fairness, maximin share (MMS), always exists and can be computed efficiently for any mixed instance with lexicographic preferences.


\subsubsection*{\myemph{\textbf{Proportionality up to one item (PROP1):}}}
In the case of additive valuations, the EF1 allocation implies the PROP1 allocation. Consequently, any algorithm that gives EF1 allocations also guarantees PROP1. \citeauthor{aziz2022fairindivisiblegoodsandchores}~\cite{aziz2022fairindivisiblegoodsandchores} proved that a \myemph{contiguous PROP1} allocation of a combination of goods and chores could be computed in polynomial time for additive utilities. This exciting result paves the way for practical algorithms that can achieve fairness and connectivity in allocating indivisible goods and chores.

\begin{tcolorbox}[enhanced,drop fuzzy shadow southwest, colframe=red!20!white!30!black,colback=LimeGreen!10]
\begin{algorithm}[H]
\caption{\myemph{EF1 for Combination of Goods and Chores with General Valuations}}\label{algo:ef1combo}
\begin{algorithmic}[1]
\State Initialize $A \leftarrow (\emptyset,\emptyset,\ldots,\emptyset)$
\State // For each good
\For{$k \leftarrow 1 \textit{ to } m$}
\State $V^g \leftarrow {i \in [n] ; g \in [m] ; | v_{ig} \ge 0}$ // Agents who value the good positively
\State Choose a source agent $j$ from the envy graph $G^g_A$
\State $A_j \leftarrow A_j \cup {k}$
\While{$G_A$ has a directed cycle $C$}
\State $A \leftarrow A^C$ // Remove the cycle from the allocation
\EndWhile
\EndFor
\State // For each chore
\For{$k \leftarrow 1 \textit{ to } m$}
\If{there is no sink agent in envy graph $G_A$}
\State $C \leftarrow $ any cycle in $T_A$
\State $A \leftarrow A^C$ // Remove the cycle from the allocation
\EndIf
\State Choose a sink agent $j$ from the envy graph $G_A$
\State $A_j \leftarrow A_j \cup {k}$
\EndFor
\Return Allocation $A$
\end{algorithmic}
\end{algorithm}
\end{tcolorbox}

\subsubsection*{\myemph{Proportionality up to any item  (PROPX):}}

PROPX requires that each agent receives a utility that is at least proportional to their share, even after we add a small good allocated among the remaining agents to their bundle or remove the smallest chore (least negative) from the agent's bundle as mentioned in Definition~\ref{def:prop}. However, PROPX may not exist even for three agents when items are only goods \citep{aziz2020polynomialPROP1PO}. Therefore, it can be concluded that PROPX may not exist in the case of a combination of goods and chores. However, this is a much less explored domain. Although we know that PROPm exists in polynomial time for goods and PROPX exists in polynomial time for chores, we have yet to explore a stronger notion than PROP1 but weaker for proportionality in the case of goods and chores.

\subsubsection*{\myemph{\textbf{Maximin Share (MMS):}}}

When allocating indivisible goods or chores among a group of agents, MMS allocation is often used to ensure fairness. However, the MMS allocation may not exist when goods or chores are indivisible. Hence, in the case of the combination of goods and chores, MMS allocation may not exist.

To address this issue, researchers have explored the concept of approximate MMS ($\alpha$-MMS) allocation, which aims to allocate goods or chores to approximate the ideal MMS allocation. The best-known approximate solution for $\alpha$-MMS in goods is $3/4+1/12n$, where $n$ is the number of agents.

In a recent study, \citeauthor{kulkarni2021indivisible} introduced the problem of finding the (near) best $\alpha \in (0,1]$ for which an $\alpha$-MMS allocation exists. They showed that an $\alpha$-MMS allocation \myemph{may not always exist} for any given $\alpha>0$, making it challenging to solve the problem for a fixed $\alpha$. To tackle this challenge, the authors developed an efficient algorithm to find an $\alpha$-MMS and Pareto Optimal (PO) allocation for the maximum $\alpha \in (0,1]$ for which it exists. This approach can provide a fair allocation of indivisible goods or chores, even when an exact MMS solution is not possible.


\subsection{Fair and Efficient Allocation}
\subsubsection*{\myemph{\textbf{Efficient EF1 allocations}}}

The concept of allocating goods and chores fairly and efficiently has been studied recently. In Section~\ref{sec:chores}, it was noted that when valuations are negative, it is still an unsolved issue to find an allocation that meets both EF1 (envy-freeness up to one item) and PO (proportional) properties. While there is a pseudo-polynomial time algorithm for determining PO and EF1 allocation in the case of goods, the existence and complexity of such an allocation for a combination of goods and chores remain unclear. However, \citeauthor{aleksandrov2019greedy} has explored a weaker form of EF1 and PO for mixed sets of goods and chores.

Another area of interest is maximizing utilitarian welfare while ensuring EF1 allocation in mixed good and chore allocation. Unfortunately, this problem has not been fully explored yet.

\subsubsection*{\myemph{\textbf{Efficient PROP1 allocations}}}
The paper by \citeauthor{aziz2020polynomialPROP1PO}~\cite{aziz2020polynomialPROP1PO} presents a \myemph{strongly polynomial time} algorithm for computing allocations that are Pareto optimal and PROP1 for both goods and chores. They demonstrate that an fPO and PROP1 allocation always exists, even for mixed utilities and any number of agents. To achieve this, they first compute a proportional fPO allocation with divisible items and then round it to preserve the fPO property and ensure PROP1. The paper by \citeauthor{branzei2019choresPROP1PO} and \citeauthor{barman2019PROP1POGoods} use the same methodology but rely heavily on the concept of \myemph{competitive equilibrium with equal incomes (CEEI)}, which does not apply to economies with a mixture of goods and chores. Therefore, the authors of \cite{aziz2020polynomialPROP1PO} construct a rounding procedure that does not rely on equilibrium prices and applies to any fPO proportional allocation of divisible items with an acyclic consumption graph. This is achieved by starting with an equal division and then finding a Pareto-dominating allocation through sequential cyclic trades, ensuring the consumption graph remains acyclic. The algorithm picks an agent who shares some items with other agents and rounds all their fractions to their advantage, breaking all other partial shares in the corresponding sub-tree. The acyclicity of the tree ensures that the algorithm terminates and returns a PROP1 allocation.

In conclusion, despite the challenges posed by the combination of goods and chores, there are fair allocation criteria that can be computed efficiently. However, there is still much room for further research, particularly in maximizing welfare while ensuring fairness constraints. We summarize the results of goods and chores in the following image.

 \begin{figure}[h]
     \centering
    \includegraphics[width=\textwidth]{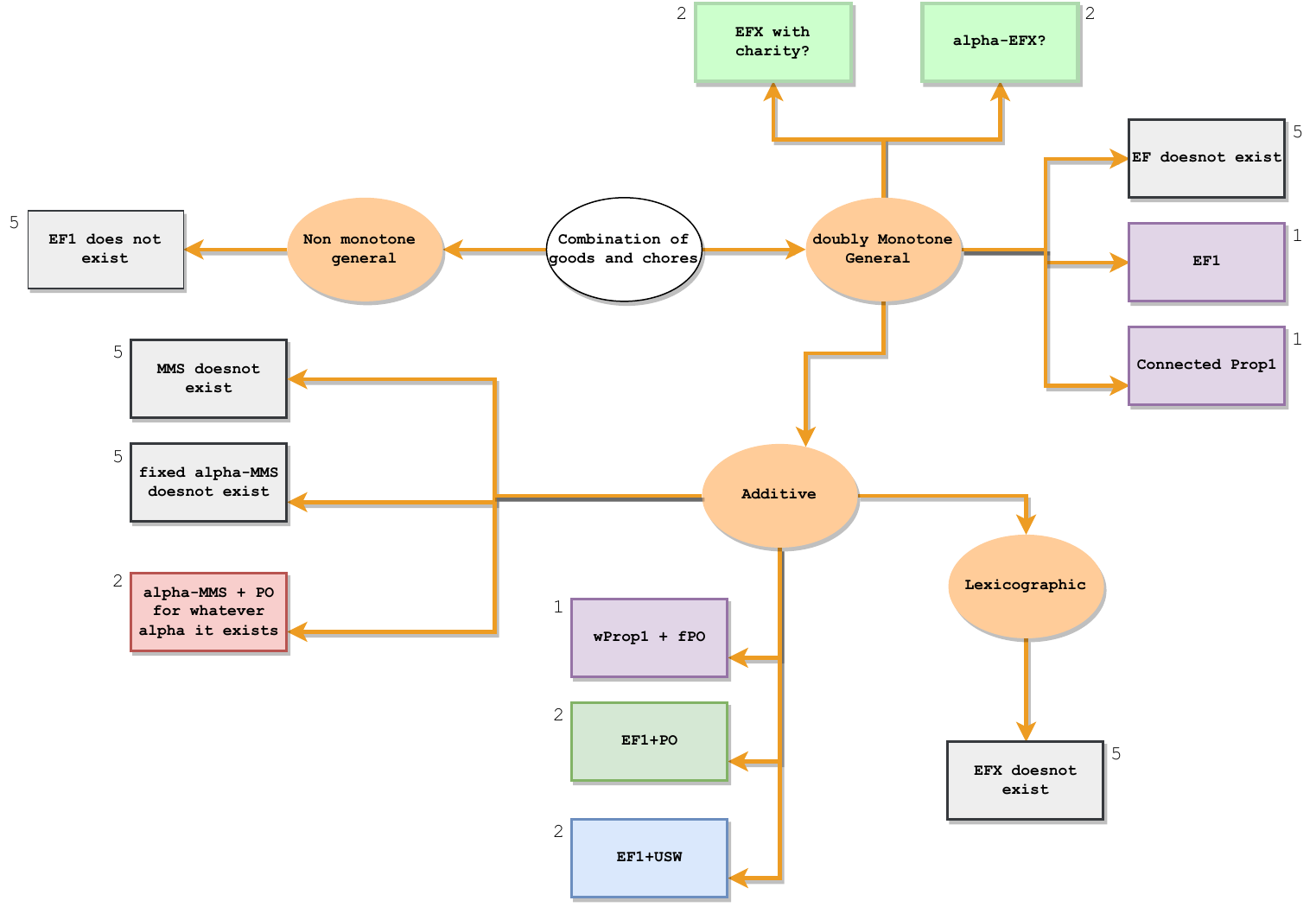}
     \caption{Summary of fairness and efficient notion for indivisible goods and chores. The Violet box (number 1) represents the polynomial-time algorithm. The blue box (number 2) represents exponential time. The red box (number 3) represents Pseudo-polynomial time. The green box (number 4) represents an open question. The grey box (number 5) represents such an allocation does not exist.}
     \label{fig:combsummary}
 \end{figure}


\section{Mixed Indivisible and Divisible Goods and Chores }
\label{sec:mixedmodel}


The world of fair division literature has long been fixated on the concept of items being either divisible or indivisible. But what happens when reality throws us a curveball, and we're suddenly faced with the challenge of allocating a mixed bag of divisible and indivisible items? As it turns out, this is a very real predicament many of us encounter daily, as evidenced by the classic example of room sharing and rent division. This scenario poses the intriguing possibility of some items being allocated fractionally while others must be claimed in their entirety. To break it down even further, let $D={k\in M\mid k \mbox{ is divisible item }}$ and $I={k\in M\mid k \mbox{ is indivisible item }}$, with $D\cap I= \emptyset \mbox{ and } D \cup I = M$. So the question becomes: how do we navigate this complex terrain of mixed divisible and indivisible items to achieve a fair and equitable division for all involved?

The study of fair division with mixed types of resources was initiated by \citeauthor{bei2021fairmixeddivisibleandindivisiblegoods} in their research, in which they explored envy-freeness for mixed goods, including both indivisible and divisible items. Further, \citeauthor{bhaskar2020approximateefforindivisiblechores} delved into envy-freeness for mixed goods/chores, encompassing both indivisible goods/chores and divisible chores. While they also attempted to investigate envy-freeness for indivisible chores and divisible goods, their research yielded limited positive results, leaving room for further exploration.

Since envy-freeness cannot be achieved for indivisible items, relaxation techniques must be considered. However, EF1 (envy-freeness up to one item) can be too weak when only divisible resources are present. This raises the question of whether we can combine a relaxation notion that captures EF and EF1. For instance, imagine Agastya and Noorie are tasked with cleaning their house (a divisible chore) and dividing a bicycle between them (an indivisible good). What would be a fair allocation in this situation? If we use EF1 for indivisible items and EF for divisible items, both would clean the house equally, and Agastya would get the bicycle. However, is this allocation truly fair? A fairer allocation would be for Agastya to clean the house and receive the bicycle. This highlights the need to reconsider our notions of fairness and efficiency in such settings.

\citeauthor{bei2021fairmixeddivisibleandindivisiblegoods} proposed a new concept called \myemph{EFM (envy-freeness for mixed goods)}, which applies to both indivisible and divisible goods. This notion was then extended by \citeauthor{bhaskar2020approximateefforindivisiblechores} to include indivisible goods/chores and divisible goods/chores. In essence, EFM means that any agent who owns divisible goods should not be envied, any agent who owns divisible chores should not envy anyone else, and, subject to these conditions, any pairwise envy should be EF1. In summary, EFM provides a more comprehensive and nuanced approach to achieving fairness in allocating mixed types of resources. Next, we define EFM for various scenarios.

\paragraph*{\myemph{Divisible Good, Indivisible Good ($D,I$ both consists of goods)\cite{bei2021fairmixeddivisibleandindivisiblegoods}}}
In the case of divisible and indivisible goods, an allocation is considered EFM if an agent's allocation contains a divisible good, then other agents compare their allocation to that agent using the EF criterion. However, if the allocation contains only indivisible goods, then the other agents compare their allocation to that agent's bundle using EF1. 
More formally, $\forall i,j \in N$,
\begin{align}
\begin{rcases}
 v_i(A_i) \ge v_i(A_j) &; \mbox{ $\exists k \in A_j$ s.t. $k$ is divisible good}\\ 
 v_i(A_i) \ge v_i(A_j \setminus k) &; \mbox{ $\exists k \in A_j; A_j \subseteq I$} 
\end{rcases}  \mbox{\myemph{ is EFM}}  \nonumber
\end{align}
\paragraph*{\myemph{Divisible Good, Indivisible Chore($D$ consists of goods, $I$ consists of chores)\cite{bhaskar2020approximateefforindivisiblechores}}}
In the case of divisible goods and indivisible chores, an allocation is EFM if an agent's allocation contains a divisible good, then other agents compare their allocation to that agent using the EF criterion. However, if the allocation contains any indivisible chore, the agent compares their allocation to other agents' allocation using EF1. 
More formally, $\forall i,j \in N$,
\begin{align}
\begin{rcases}
 v_i(A_i) \ge v_i(A_j) &; \mbox{ $\exists k \in A_j$ s.t. $k$ is divisible good}\\ 
 v_i(A_i \setminus k) \ge v_i(A_j) &; \mbox{ $\exists k \in A_i; v_{ik}<0$} 
\end{rcases}  \mbox{\myemph{ is EFM}}  \nonumber
\end{align}
\paragraph*{\myemph{Divisible Chore, Indivisible Good($D$ consists of goods, $I$ consists of chores)\cite{bhaskar2020approximateefforindivisiblechores}}}
In the case of divisible chores and indivisible goods, an allocation is EFM if an agent's allocation contains a divisible chore, then the agent compares their allocation to other agents' allocation using the EF criterion. However, if the allocation contains an indivisible good, the other agents compare their allocation to that agent using EF1
More formally, $\forall i,j \in N$,
\begin{align}
\begin{rcases}
 v_i(A_i) \ge v_i(A_j) &; \mbox{ $\exists k \in A_i$ s.t. $k$ is divisible chore} \; \mbox{or} \\ 
 v_i(A_i) \ge v_i(A_j \setminus k) &; \mbox{ $\exists k \in A_j; v_{ik}>0$}
\end{rcases}  \mbox{\myemph{ is EFM}}  \nonumber
\end{align}
\paragraph*{\myemph{Divisible Chore, Indivisible Chore ($D$ consists of chores, $I$ consists of chores) \cite{bhaskar2020approximateefforindivisiblechores}}}
Lastly, in the case of divisible and indivisible chores, an allocation is EFM if an agent's allocation contains a divisible chore, then other agents compare their allocation to that agent using the EF criterion. However, if the allocation contains an indivisible chore, the agent compares their allocation to other agents' allocation using EF1. 
More formally, $\forall i,j \in N$,
\begin{align}
\begin{rcases}
 v_i(A_i) \ge v_i(A_j) &; \mbox{ $\exists k \in A_i$ s.t. $k$ is divisible chore} \; \mbox{or} \\ 
 v_i(A_i \setminus k) \ge v_i(A_j ) &; \mbox{ $\exists k \in A_i; A_i\subseteq I$}
\end{rcases}  \mbox{\myemph{ is EFM}}  \nonumber
\end{align}

The paper by \citeauthor{bei2021fairmixeddivisibleandindivisiblegoods}\cite{bei2021fairmixeddivisibleandindivisiblegoods} examines the problem of fairly dividing a set of heterogeneous divisible goods and a collection of indivisible goods. The authors prove that an allocation that satisfies envy-freeness for mixed resources (EFM) for mixed goods with additive valuations for divisible goods and general monotone indivisible goods always exists. The algorithm used to achieve this requires an oracle for computing a perfect allocation in cake cutting but can compute an EFM allocation in a polynomial number of steps. The algorithm starts by obtaining an arbitrary EF1 allocation (using the \myemph{Envy-cycle elimination} algorithm) of the indivisible goods, and then allocates the divisible goods to obtain an EFM allocation. While the paper's algorithm is only proven for additive valuation for indivisible goods, \cite{bhaskar2020approximateefforindivisiblechores} shows that it also works for general monotone valuation.

Additionally, the authors present two algorithms for computing an EFM allocation for special cases without using the perfect allocation oracle. The first algorithm is for two agents with general additive valuations in the Robertson-Webb model, and the second is for any number of agents with piece-wise linear valuation functions.

In the Robertson-Webb model, the authors present an algorithm that computes an \myemph{$\epsilon$-EFM allocation} with a polynomial running time complexity that is dependent on the number of agents, the number of indivisible goods, and $1/\epsilon$. The authors note that it is unclear whether an EFM allocation can be computed in a finite number of steps in the Robertson-Webb model in general. The algorithm does not require a perfect allocation oracle, which is an appealing result due to its polynomial running time complexity. In contrast, a bounded exact EFM protocol, if it exists, is likely to require a large number of queries and cuts. For instance, in the special case when all resources are divisible, EFM reduces to EF in cake cutting, and the best-known protocol \cite{AzizandMackenzie} for this problem has a very high query complexity (a tower of exponents of $n$).

\citeauthor{bhaskar2020approximateefforindivisiblechores}\cite{bhaskar2020approximateefforindivisiblechores} investigates a mixed model consisting of indivisible items (goods/chores) and a divisible heterogeneous chore. The authors extend the definition of EFM to this model and prove that an EFM allocation always exists. They use a similar algorithmic approach to that in \cite{bei2021fairmixeddivisibleandindivisiblegoods}, starting with an arbitrary EF1 allocation of the indivisible item and then allocating the cake to obtain an EFM allocation.

Finally, the authors of \cite{bhaskar2020approximateefforindivisiblechores} also study a mixed model with \myemph{indivisible chores and divisible goods}, which proves to be more challenging. Unlike previous cases, it is not possible to start with an arbitrary EF1 allocation of the indivisible items and allocate the cake to obtain an EFM allocation. However, the authors show that an EFM allocation exists for two special cases in this model: when each agent has the same ranking over the chores, and when the number of chores is at most one more than the number of agents. In particular, for two agents, an EFM allocation always exists in this setting, and with $n$ agents, $m$ additive indivisible chores, and cake, where $m \le n+1$, an EFM allocation exists.

We have summarized the results for a mixed setting of divisible and indivisible item instances in Table~\ref{tab:mixedmodelresults}. From the table, it can be observed that this domain has been relatively less explored but is highly relevant and has the potential to yield exciting findings.





\begin{table}[h]
\centering
\caption{\label{tab:mixedmodelresults}Summary of Fair Division for Mixed Model}
\resizebox{16cm}{!}{
\begin{tabular}{|c|c|c|c|c|c|c|c|c|}
\hline
\rowcolor[HTML]{b7df86}
{Fairness}& {Agents}  &   {\begin{tabular}[c]{@{}l@{}}Divisible\\ Item\end{tabular}
} & {\begin{tabular}[c]{@{}l@{}}Divisible\\ Valuation\end{tabular}
}  & {\begin{tabular}[c]{@{}l@{}} Indivisible\\ Item\end{tabular}
} & {\begin{tabular}[c]{@{}l@{}}Indivisible\\ Valuation\end{tabular}
} &  {Paper} & {Computation} & {Details} \\ \hline 
EFM
& any &good  &  additive &   good &  general & \citeauthor{bei2021fairmixeddivisibleandindivisiblegoods}~\cite{bei2021fairmixeddivisibleandindivisiblegoods} & polynomial & With perfect allocation oracle \\ 
\rowcolor[HTML]{dbefc3}
EFM & any  & good  & additive & chores & general  &? & ? &  \\ \hline
\multirow{3}{*}{EFM} & \multirow{3}{*}{any}  &  \multirow{3}{*}{chores} & \multirow{3}{*}{additive}  & {good} & \multirow{3}{*}{general}  & \multirow{3}{*}{\citeauthor{bhaskar2020approximateefforindivisiblechores}~\cite{bhaskar2020approximateefforindivisiblechores}}  & \multirow{3}{*}{polynomial} & \multirow{3}{*}{With perfect allocation orcale}  \\ \cline{5-5}
  &  &    &  & {chores} &  &  &  &  \\ \cline{5-5}
  &  &  &  & {combination} &  &  &  &  \\ \hline
  &  &   \multicolumn{1}{l|}{good} &  & \multicolumn{1}{l|}{good} & general & \citeauthor{bei2021fairmixeddivisibleandindivisiblegoods}~\cite{bei2021fairmixeddivisibleandindivisiblegoods} & polynomial &  \\
  \cline{3-3} \cline{5-8}
  \rowcolor[HTML]{dbefc3}
  &   & \multicolumn{1}{l|}{good} &  & \multicolumn{1}{l|}{} &  &  &  &  \\ \cline{3-3}
  \rowcolor[HTML]{dbefc3}
 \multirow{-3}{*}{EFM} & \multirow{-3}{*}{$n=2$}  & \multicolumn{1}{l|}{chore} & \multirow{-3}{*}{general} & \multicolumn{1}{l|}{\multirow{-2}{*}{combination}} & \multirow{-2}{*}{general} & \multirow{-2}{*}{?} & \multirow{-2}{*}{?} & \multirow{-3}{*}{With   perfect allocation orcale} \\ \hline
  &  &    &  & {good} &  &  & polynomial &  \\ \cline{5-5} \cline{7-8}
   \rowcolor[HTML]{dbefc3}
  &  &   {\multirow{-2}{*}{good}} &  &  &  & ? & ? &  \\ \cline{3-3} \cline{7-8}
   \rowcolor[HTML]{dbefc3}
 \multirow{-3}{*}{EFM} & \multirow{-3}{*}{any}  & {chore} & \multirow{-3}{*}{piece   wise linear} & {\multirow{-2}{*}{combination}} & \multirow{-3}{*}{general} & ? & ? & \multirow{-3}{*}{With   perfect allocation orcale} \\ \hline
 
  & $n=2$ &  {good} & additive &  &  &  &  &  \\ \cline{2-4} 
  & any & {good} & additive identical for n-1 &  &  &  &  &  \\ \cline{2-4} 
 \multirow{-3}{*}{EFM} &  $m\le n+1$ & {good} & additive & {\multirow{-3}{*}{chore}} & \multirow{-3}{*}{general} & \multirow{-3}{*}{\citeauthor{bhaskar2020approximateefforindivisiblechores}~\cite{bhaskar2020approximateefforindivisiblechores}} & \multirow{-3}{*}{polynomial} & \multirow{-3}{*}{With   perfect allocation orcale} \\ \hline
  &  &   &  & {good} &  & \citeauthor{bei2021fairmixeddivisibleandindivisiblegoods}~\cite{bei2021fairmixeddivisibleandindivisiblegoods} & polynomial &  \\ 
  \cline{5-5} \cline{7-8}
     \rowcolor[HTML]{dbefc3}
  &  &   {\multirow{-2}{*}{good}} &  &  &  &  &  &  \\ \cline{3-3}
     \rowcolor[HTML]{dbefc3}
  \multirow{-3}{*}{$\epsilon$-EFM} & \multirow{-3}{*}{any} &  {chores} & \multirow{-3}{*}{additive} & {\multirow{-2}{*}{combination}} & \multirow{-3}{*}{general} & \multirow{-2}{*}{?} & \multirow{-2}{*}{?} & \multirow{-3}{*}{Without perfect allocation orcale} \\ \hline
\end{tabular}}

\end{table}

\section{Summary of Complexity Results for Fair Division}
\label{sec:otherresults}
The problems of computing allocations that maximize the egalitarian welfare (utility of the worst-off agent) or Nash welfare (geometric mean of the utilities of agents) have been shown to be NP-hard, as demonstrated in prior research by~\citeauthor{nguyen2014minimizing}. Moreover, even computing an exact maximin share for an agent with additive valuations has been proven to be an NP-complete problem, as shown through a reduction from the NP-hard problem of Partition. Additionally, in the case of agents with binary additive valuations for goods, checking for the existence of an envy-free (EF) allocation has also been shown to be NP-complete, as demonstrated in studies by~\citeauthor{aziz2015fairordinal,hosseini2020fair}. Furthermore, the problem of making efficiency improvements in fair allocation has been identified as NP-hard in research by~\citeauthor{aziz2019constrained}, further highlighting the computational complexity of these allocation problems.

In addition to the computational complexity, it is important to consider the likelihood of finding a fair allocation in real-world scenarios. Research has explored the conditions under which a fair allocation is likely to exist or not, shedding light on the practical feasibility of achieving fairness in allocation mechanisms.

\section{Other Important Results}
In this section, we will provide a brief overview of other important aspects in the field of fair division, including the likelihood of finding a fair allocation, and the price of fairness.

\subsection{Likelihood of Finding a Fair Allocation}
\citeauthor{dickerson2014computational} was the first to address this question, under the assumption that utilities are additive and each agent's utilities for individual items are drawn from probability distributions. They established that an envy-free allocation is likely to exist with high probability when the number of items $m$ is at least $\Omega(n log n)$, but not when $m = n + o(n)$. \citeauthor{suksompong2016asymptotic} investigated the asymptotic existence of proportional allocations - which are weaker than envy-free allocations under the additivity assumption - and showed that such allocations occur with high probability provided that either $m$ is a multiple of $n$ or $m = w(n)$.

\citeauthor{manurangsi2021closing} showed that EFX allocations, which satisfy envy-freeness up to any item, exist with high probability for any number of agents and items under the assumption that the valuations of the agents are drawn at random from a probability distribution. When agents' valuations for individual items are randomly drawn from a probability distribution, they show that the classical round-robin algorithm is likely to result in an envy-free allocation, particularly when the number of items $m$, satisfies the condition $m = \Omega(n \log n/ \log \log n)$. Furthermore, they demonstrated that a proportional allocation is likely to exist with high probability as long as the number of items, $m$, is greater than or equal to the number of agents, $n$. Additionally, an allocation that satisfies envy-freeness up to any item (EFX) is likely to be present for any relationship between $m$ and $n$. They also demonstrated the challenge of showing the existence of EFX allocations, as in instances with two agents there can be as few as two EFX allocations, while the number of EF1 allocations is always exponential in the number of items.

\citeauthor{Amanatidismms2017} undertook a probabilistic analysis and proved that in randomly generated instances, maximin share allocations exist with high probability, motivated by the apparent difficulty in establishing lower bounds in calculating $\alpha$-MMS. Furthermore,~\citeauthor{procaccia2014mmsdoesntexist} showed that the non-existence of MMW (maximin share with weights) allocation requires a number of items that is exponential in the number of players. In contrast, if the number of items is only slightly larger than the number of players, an MMS allocation is guaranteed to exist~\cite{bouveret2016characterizingconflictsinFD}. This raises the question of what is the largest number of items for which an MMS allocation is guaranteed to exist.

In conclusion, the likelihood of finding a fair allocation depends on various factors such as the specific fairness notion, the type of valuations, and the number of agents and items involved. Probabilistic analyses have been conducted to understand the conditions under which fair allocations are likely to exist, shedding light on the practical feasibility of achieving fairness in real-world scenarios.

\subsection{Price of fairness}
The concept of the price of fairness quantitatively measures the \myemph{tradeoff between fairness and social welfare}. It represents the ratio of the maximum welfare of any allocation to the maximum welfare of any allocation that satisfies the desired fairness notion, and it provides insights into the efficiency loss incurred in achieving fairness.

The authors,~\citeauthor{caragiannis2012efficiency}, investigated the effect of fairness on the efficiency of resource allocations. They examined three fairness criteria, namely proportionality, envy-freeness, and equitability, for allocations of divisible and indivisible goods and chores. They presented a set of findings on the price of fairness under each of these criteria that quantify the loss in efficiency in fair allocations compared to optimal ones.

For indivisible goods, the authors derived an exact bound of $n+1/n$ on the price of proportionality, while they demonstrated that the price of envy-freeness is $\theta(n)$. In contrast, for indivisible chores, they established an exact bound of $n$ on the price of proportionality, but both the price of envy-freeness and equitability are infinite. These results indicate that, in the case of indivisible chores, envy-freeness and equitability are typically incompatible with efficiency.

Their work provides insights into the trade-off between efficiency and fairness in resource allocation, considering both divisible and indivisible goods. However, a limitation of their study is that fairness notions may not always be satisfied for indivisible goods, which they addressed by ignoring such instances in their analysis of the price of fairness. This omission may not capture certain scenarios that can arise in practice. Furthermore, the assumption that envy-free allocations are always proportional may not necessarily hold true, as there are instances where proportional allocations exist but envy-free allocations do not.

In the study by \cite{bei2021price}, the focus shifts from classical fairness notions, such as envy-freeness, proportionality, and equitability, which may not always be achievable for indivisible goods, to notions with guaranteed existence, including envy-freeness up to one good (EF1), balancedness, maximum Nash welfare (MNW), and leximin. They also introduce the concept of the strong price of fairness, which captures the efficiency loss in the worst fair allocation. They provide tight or asymptotically tight bounds on the worst-case efficiency loss for allocations satisfying these notions, for both the price of fairness and the strong price of fairness. 

Specifically, the study shows that for the price of EF1, a lower bound of $\Omega(\sqrt{n})$ and an upper bound of $O(n)$ are provided. Additionally, it is shown that two common methods for obtaining an EF1 allocation, namely the round-robin algorithm and MNW, have a price of fairness of linear order (with the exact price being n for round-robin), indicating that these methods cannot improve the upper bound for EF1. For MNW, MEW, and leximin, an asymptotically tight bound of $\theta(n)$ on the price of fairness is proven. The results suggest that round-robin is a promising allocation method, as it produces an EF1 allocation with high welfare, is simple and intuitive, and always produces a balanced allocation.

The authors of \cite{barman2020optimal} address the impact of fairness guarantees on social welfare in the allocation of indivisible goods by resolving the price of two well-known fairness notions: envy-freeness up to one good (EF1) and approximate maximin share (MMS). They show that the price of fairness is $O(\sqrt{n})$ for both EF1 and 1/2-MMS, using different techniques. The $\Omega(\sqrt{n})$ lower bound due to~\citeauthor{bei2021price} is matched by their upper bound, which holds for the more general class of subadditive valuations, as opposed to additive valuations. Their work therefore resolves this open question for all valuation classes between additive and subadditive and also settles the price of proportionality up to one good (Prop1) as $\theta(\sqrt{n})$ for additive valuations.

For the 1/2-MMS fairness notion and additive valuations, the authors also show that the price of fairness is $\theta(\sqrt{n})$ using a different algorithm. They prove that for a fixed $\epsilon>0$, a $(1/2-\epsilon)$-MMS allocation with welfare within $O(\sqrt{n})$ factor of the optimal can be computed in polynomial time. Overall, their main contribution is comprehensively settling the price of EF1 and 1/2-MMS fairness notions, and providing efficient algorithms for obtaining their upper bounds.

\citeauthor{li2021wpropxchores} investigate the price of fairness (PoF), which measures the loss in social welfare when enforcing allocations to be (weighted) proportional (PROPX). They prove that the tight ratio for PoF is $\theta(n)$ for symmetric agents, indicating that the loss in social welfare can be significant. However, they also show that the PoF can be unbounded for asymmetric agents, suggesting that fairness constraints can have a drastic impact on social welfare in such cases.

\subsection{Different Settings in Fair Division}
After exploring different notions of fairness and their complexities, we will now provide a brief overview of various settings to highlight the intricacies of fair division. It's important to note that fairness is subjective and can vary depending on the setting. What may be considered fair in one context may not be viewed as fair in another. Therefore, it is necessary to outline all the possible settings before delving into a detailed analysis of a specific setting.
\smallskip

\noindent\myemph{\textbf{Different Types of Items}}
The literature on fair division encompasses various items, including divisible, indivisible, or a combination of both. Divisible items can be divided into smaller portions, such as disk space in a CPU, cake for children, money for employees, land among siblings, or rent among housemates. Divisible items can be either homogeneous, where the value is uniform across the item or heterogeneous, where the values may differ. There is rich literature on ensuring envy-free allocation for divisible items, with numerous possibilities to explore.

On the other hand, indivisible items cannot be further divided and must be allocated entirely to one person or not at all. Indivisible items include books, cars, antiques, laptops, and trees. Ensuring fairness in allocating indivisible items is more challenging than divisible items. For instance, allocating one car among two people can never be truly fair, as giving it to one person would inherently be unfair to the other. Achieving fairness in allocating indivisible items often requires considering both efficiency and fairness measures. Indivisible items can also include goods or chores, such as distributing free books based on people's preferences, whereas allocating a book outside someone's preference can result in negative utility.

In practice, situations often arise where a mixed set of divisible and indivisible items must be allocated, for example, dividing rent and room allocation among friends sharing an apartment with different room preferences or dividing chores and a mobile phone among roommates. Allocating fairly in such mixed settings requires re-considering fairness and efficiency to ensure a more equitable allocation.

\smallskip
Next, we list various possibilities for agents to state their valuation/preferences over a bundle of items.

\noindent\myemph{\textbf{Different Types of Valuation functions}}

In this section, we consider different ways agents can express their preferences or valuations for bundles of items. First, in the combinatorial setting, each agent is asked to value $2^m$ possible bundles of $m$ indivisible items that must be divided among $n$ agents. Various types of valuations can be used for such bundles, such as sub-additive, super-additive, and xor. As an illustration, consider the items of a left shoe and a right shoe, which are examples of super-additive items due to their complementarity. This means that the combined value of the left and right shoes is greater than the sum of their individual values. On the other hand, items such as the Luxor Pen and the Pentonic Pen are sub-additive, as they are substitutes for each other. Agents would typically only desire one pen, thus, the value of each individual pen is either greater than or equal to the combined value of both pens. However, it can be challenging for agents and social planners to evaluate such a large number of valuations, even with learning-based preference elicitation techniques. In the case where 30 items are to be distributed, the system is requesting agents to evaluate over a billion item-bundle values, which can be an exceedingly complex task. There exists a plethora of literature on the intricacies involved in eliciting preferences~\cite{sandholm2006preference}.

Second, the literature mainly assumes that agents have additive valuations over bundles of items, which capture the essence of agents' valuations in a simple yet accurate manner. Additionally, many real-world scenarios involve agents who only express their preferences over the items they like or dislike, known as binary valuations, and there are several modifications to this class, such as dichotomous preferences, (0,1)-OXS, matroid rank valuation functions, single-mindedness, etc.

Third, agents may only report partial knowledge of their valuations, such as their top $k$ items out of a large list of items, due to the difficulty of evaluating additive valuations for many items. This is known as the partial information setting. This scenario arises when thousands of items are to be divided among a group of interested agents, making it challenging to elicit additive valuation. Hence agents report only their top favorite items. For example, in a research institute, a department such as Security may not report their valuation of an item such as a microscope and instead focus on valuing items that are more important to them. The paper~\cite{nisargshah2021fairwithpartialinformation} redefines fairness notions applicable to this setting and provides results on EF1 and MMS fairness.

In another scenario, agents only report partial knowledge about their valuations, i.e., they report the top $k$ valued item out of all the listed items. This scenario arises when thousands of items are to be divided among a group of interested agents, making it challenging to elicit additive valuation. Hence agents report only their top favorite items. For example, in a research institute, a department such as Security may not report their valuation of an item such as a microscope. Instead, they focus on value items that are more important to them. The paper~\cite{nisargshah2021fairwithpartialinformation} redefines fairness notions applicable to this setting and provides results on EF1 and MMS fairness.

Fourth, some agents may find it tedious to evaluate the exact value of a bundle of items, in which case ordinal valuations can be used. In this setting, each agent is equipped with a preference relation that expresses their preference for one set of items over another. Ordinal preferences are often used in practical applications where the algorithm cannot collect complete information on agent preferences. In this case, the algorithm can use partial information to compute approximately fair allocations~\cite{aziz2015fairordinal,bouveret2010fairordinalECAI}. The literature has studied various notions of envy-freeness and proportionality in the ordinal setting. For example, people ranking their preferences in voting, schools, or university applications can be modeled using ordinal preferences. Finally, recent studies have shown that approximately fair allocations can be computed using ordinal preferences, even with partial information.

Lastly, Researches are also exploring the problem of allocating indivisible goods among strategic agents and focus on settings where monetary transfers are not available. Truthfulness is a key desideratum in this context, meaning mechanisms should ensure that participating agents cannot gain by misreporting their valuations. However, achieving the central objectives of fairness, economic efficiency, and truthfulness together in settings without monetary transfers is challenging, even under additive valuations~\cite{barman2021truthful}.

\smallskip
Next, we list various possibilities in different settings arrive to model real-life settings introducing complexity in allocating items fairly.

\noindent\myemph{\textbf{Different Types of Setting}}

\myemph{\textbf{Online setting}}: Most of the literature on fair division deals with a fixed number of agents and items, which is not reflective of real-life settings where the number of agents or items may change over time~\cite{aleksandrov2020onlinesurvey,walsh2011online}. Examples of such settings include allocating parking slots to cars, donations to changing recipients, allocating resources to system processes, and enrolling students in Ph.D. programs. Papers have been written on online fair division in various settings, including fixed agents and changing items, changing agents and fixed items, and changing agents and items. Traditional offline fairness approaches may not be effective in these online settings.

\myemph{\textbf{Two-sided matching settings}}: Many real-life scenarios involve two-sided fairness, where fairness needs to be ensured for both parties involved~\cite{aziz2020developments}. Examples include online platforms like Amazon and Airbnb, matching students to schools or employees to employers, and allocating advisors to students in Ph.D. programs. Papers have been written on recent developments and guarantees for two-sided fairness, such as double envy-freeness up to one match (DEF1) and double maximin share guarantees (DMMS), which may not exist in certain cases.

\myemph{\textbf{General models with arbitrary shares}}: There is a growing literature on fair division where agents may have arbitrary and possibly unequal shares of items~\cite{payan2021will}. Examples include peer review in academia, where papers and reviewers have preferences for each other, and conference or course allocation settings, where conflicting items need to be fairly allocated. Papers have been written on these settings and additional constraints must be considered.

\myemph{\textbf{Fair division with unequal entitlements}}: In specific scenarios, agents may have unequal entitlements to items based on their membership status or employment type, and fairness needs to be ensured accordingly~\cite{fairallocationofconglictingitems}. Examples include membership plans in a resort, where different membership levels have different entitlements, and resource allocation for full-time and part-time employees in a company. Papers have been written on the fair allocation of items with unequal entitlements and how fairness notions can be applied in such settings.

\myemph{\textbf{Public Goods}}:
In fair resource allocation, most work focuses on private goods that can only be assigned to a single agent. However, some goods are public and can be enjoyed by multiple agents simultaneously. Public goods allocation problems can arise in various contexts, such as voting for the allocation of public resources like schools, libraries, and museums in a city. Naively tallying votes may not always result in a fair allocation of public goods, as it may ignore the preferences of a large minority. Fairness axioms like proportionality do not capture the idea of fairness in public goods allocation, as they focus on individual agents' perspectives. Public goods allocation problems require striking a balance between agents with different preferences, as opposed to private goods allocation where tension arises between agents with similar preferences. The literature has moved beyond the private goods model to study fair allocation for public goods in various applications such as participatory budgeting, committee selection, and shared memory allocation. Proportional fairness, which balances fairness and efficiency, is a commonly studied notion in allocating public goods. One limitation of the current literature is that it mainly considers "one-shot" settings where all agents' preferences are known beforehand. At the same time, real-world scenarios often involve dynamic and online allocation decisions. Recent works have focused on designing online algorithms for fair resource allocation to capture the dynamic nature of fair division problems in practice~\cite{nisargshahpublicgoods18,banerjeeproportionally}
.

\myemph{\textbf{Randomized Allocation}}
In the literature survey, the concept of randomized allocation for indivisible goods is discussed. The idea of constructing a fractional assignment and implementing it as a lottery over pure assignments was introduced~\citeauthor{hylland1979efficient}. The key question addressed is whether ex-ante envy-freeness can be achieved in combination with ex-post envy-freeness up to one good. It is shown that if randomization is allowed, a simple allocation where the entire set of goods is allocated to one agent at random can achieve ex-ante envy-freeness, but induces a large amount of envy ex-post. This has led to significant research on fairness in deterministic allocations of indivisible goods, such as envy-freeness up to one good (EF1), which requires removing envy by eliminating at most one good from the envied agent's bundle. Lotteries and randomization are commonly used to break ties in allocating indivisible objects, where agents may have equal priority rights for some objects. Examples of such allocation problems include public housing associations assigning apartments to residents, school districts assigning seats to students, and childcare cooperatives assigning chores to members. Furthermore, the concept of achieving ex-ante and ex-post envy-freeness through randomized allocations is discussed. It is shown that achieving ex-ante envy-freeness does not guarantee fairness in ex-post envy, and different allocations may have different levels of fairness~\cite{freeman2020Exante,aziz2020ExAnteWine}. 

\myemph{\textbf{Repeated Allocation}} The literature explores various allocation mechanisms to achieve fairness in the repeated allocation of indivisible goods without the use of monetary transfers. This is relevant in scenarios such as allocating server time, rooms to students, or food among food banks. Additionally, the problem of fair matching has been studied in traditional setups such as marriage matching or matching medical residents to hospitals. Still, modern online matching platforms require investigating the two-sided fairness of providers and customers in a platform performing repeated matchings of providers and customers over time. The fair distribution of income for providers in on-demand and marketplace platforms is crucial for ensuring long-term stable platform operation, and its study has received little attention so far despite the increasing dependence of people on the sharing economy to earn a living~\cite{repeatedmatchingpaper19,gorokh2019remarkableRepeated}.

\myemph{\textbf{Group Fairness}} In the literature, group fairness is explored in the context of resource allocation in settings such as corporate environments~\cite{conitzer2019groupfairness}. For example, a manager may need to allocate resources such as interns, conference rooms, or time slots for shared machines to employees. The manager may aim to ensure that no business team envies another team and that there is no envy between genders, locations, roles, and other groups. Envy-freeness alone may not be sufficient in this setting, as an allocation that is envy-free up to one good may still result in significant levels of inequality and envy between groups of players. Another example can be, consider a university that needs to allocate research funding to its faculty members. The university may want to ensure that no department envies another department, that junior faculty do not envy senior faculty, that female faculty do not envy male faculty, that faculty members from one research area do not envy those from another area, and so on. Simply ensuring envy-freeness may not be sufficient in this case, as it may still result in significant levels of inequality and envy between different groups of faculty members. Therefore, the university may need to explore alternative fairness criteria and allocation mechanisms to ensure a fair and efficient allocation of research funding.

\myemph{\textbf{Externalities}} The concept of externalities in allocation refers to situations where an agent's utility not only depends on their own bundle but also on the bundles allocated to other agents~\cite{mishra2022fair,externalitiesdivisbleijcai13,externalitiesghodsi}. This scenario arises frequently in the allocation of essential goods such as hospital beds, ventilators, and vaccines during emergencies like the COVID-19 pandemic. The allocation of critical assets to one group in the division of assets among conflicting groups can also cause negative externalities for the other group's functionality. On the other hand, certain goods generate positive consumption externalities, such as the increased value of a PlayStation to an agent if more of their friends also own one. Positive externalities are also observed in healthcare, where individuals who are vaccinated reduce the risk of contraction for others around them.

\section{Summary}
\label{sec:summary}
In conclusion, fair resource allocation is a challenging problem that has been studied extensively in the literature. In this survey, we focused on the fair allocation of indivisible goods and chores, as well as mixed instances that include both indivisible and divisible items. We discussed various fairness definitions, including EF, EF1, EFX, PROP, PROP1, PROPX, PROPm, MMS, and $\alpha$-MMS, and their corresponding fair and efficient allocations, such as EF1+PO, EF1+USW, PROP1+PO, PROP1+USW, PROPm+PO/USW, and EFX+PO. All the existing results are summarized in Table~\ref{tab:summarytable1} and ~\ref{tab:mixedmodelresults}. We also touched on computational complexity and the likelihood of finding a fair allocation in different settings, such as different types of items, valuations, and settings. Finally, we considered the price of fairness and its impact on the efficiency of the allocation.
Overall, this survey provides a comprehensive overview of the state-of-the-art approaches to fair resource allocation for indivisible goods and chores. It also highlights the need for further research in this area, especially in developing efficient algorithms that can handle more complex settings and achieving a better trade-off between fairness and efficiency.
Equitably dividing resources poses numerous challenges, prompting researchers in recent literature to investigate data-driven approaches in game theory and mechanism design. 

\section*{Acknowledgement}
The authors would like to thank Siddharth Barman for pointing out technical corrections.

\printbibliography

\end{document}